\documentclass[journal,twoside,web]{ieeecolor}
\usepackage{tuffc}
\usepackage{cite}
\usepackage{amsmath,amssymb,amsfonts}
\usepackage{algorithmic}
\usepackage{graphicx}
\usepackage{textcomp}
\usepackage{multirow}
\usepackage{makecell}
\usepackage{algorithmic}
\usepackage{graphicx}
\usepackage{textcomp}
\usepackage{array}
\usepackage{booktabs}
\usepackage{wrapfig,colortbl}
\definecolor{abstractbg}{rgb}{1,0.969,0.914}
\def\BibTeX{{\rm B\kern-.05em{\sc i\kern-.025em b}\kern-.08em
    T\kern-.1667em\lower.7ex\hbox{E}\kern-.125emX}}
\begin{document}
\title{High-Quality Passive Acoustic Mapping with the Cross-Correlated Angular Spectrum Method}
\author{Yi Zeng, Hui Zhu, \IEEEmembership{Student Member, IEEE}, Jinwei Li, Jianfeng Li, Fei Li, Shukuan Lu, and Xiran Cai, \IEEEmembership{Member, IEEE}
\thanks{This work was supported by the Shanghai Sailing Program under Grant 21YF1429300. (Corresponding author: Xiran Cai and Shukuan Lu.)}
\thanks{Yi Zeng is with the School of Information Science and Technology, ShanghaiTech University, Shanghai 201210, China (e-mail: zengyi2022@shanghaitech.edu.cn).}
\thanks{Jinwei Li and Jianfeng Li are with the Gene Editing Center, School of Life Science and Technology \& State Key Laboratory of Advanced Medical Materials and Devices, ShanghaiTech University, Shanghai 201210, China (e-mail: lijw1@shanghaitech.edu.cn; lijf1@shanghaitech.edu.cn).}
\thanks{Hui Zhu is with the School of Information Science and Technology, ShanghaiTech University, Shanghai 201210, China, also with the Shanghai Advanced Research Institute, Chinese Academy of Sciences, Shanghai 201210, China, and also with the University of Chinese Academy of Sciences, Beijing 100049, China (e-mail: zhuhui@shanghaitech.edu.cn).}
\thanks{Fei Li is with the Paul C. Lauterbur Research Center for Biomedical Imaging, Shenzhen Institutes of Advanced Technology, Chinese Academy of Sciences, Shenzhen 518055, China (e-mail: fei.li@siat.ac.cn).}
\thanks{Shukuan Lu is with the Key Laboratory of Biomedical Information Engineering of Ministry of Education, Department of Biomedical Engineering, School of Life Science and Technology, Xi’an Jiaotong University, Xi’an 710049, China (e-mail: xjtusklu@mail.xjtu.edu.cn).}
\thanks{Xiran Cai is with the School of Information Science and Technology and Shanghai Engineering Research Center of Intelligent Vision and Imaging, ShanghaiTech University, Shanghai 201210, China and also with The Key Laboratory of Biomedical Imaging Science and System, Chinese Academy of Sciences, Shenzhen 518055, China (e-mail: caixr@shanghaitech.edu.cn).}}

\IEEEtitleabstractindextext{%
\fcolorbox{abstractbg}{abstractbg}{%
\begin{minipage}{\textwidth}\rightskip1em\leftskip\rightskip\bigskip
\begin{wrapfigure}[21]{r}{2.4in}%
\vspace{-1pc}\hspace{-2pc}\includegraphics[width=2.4in]{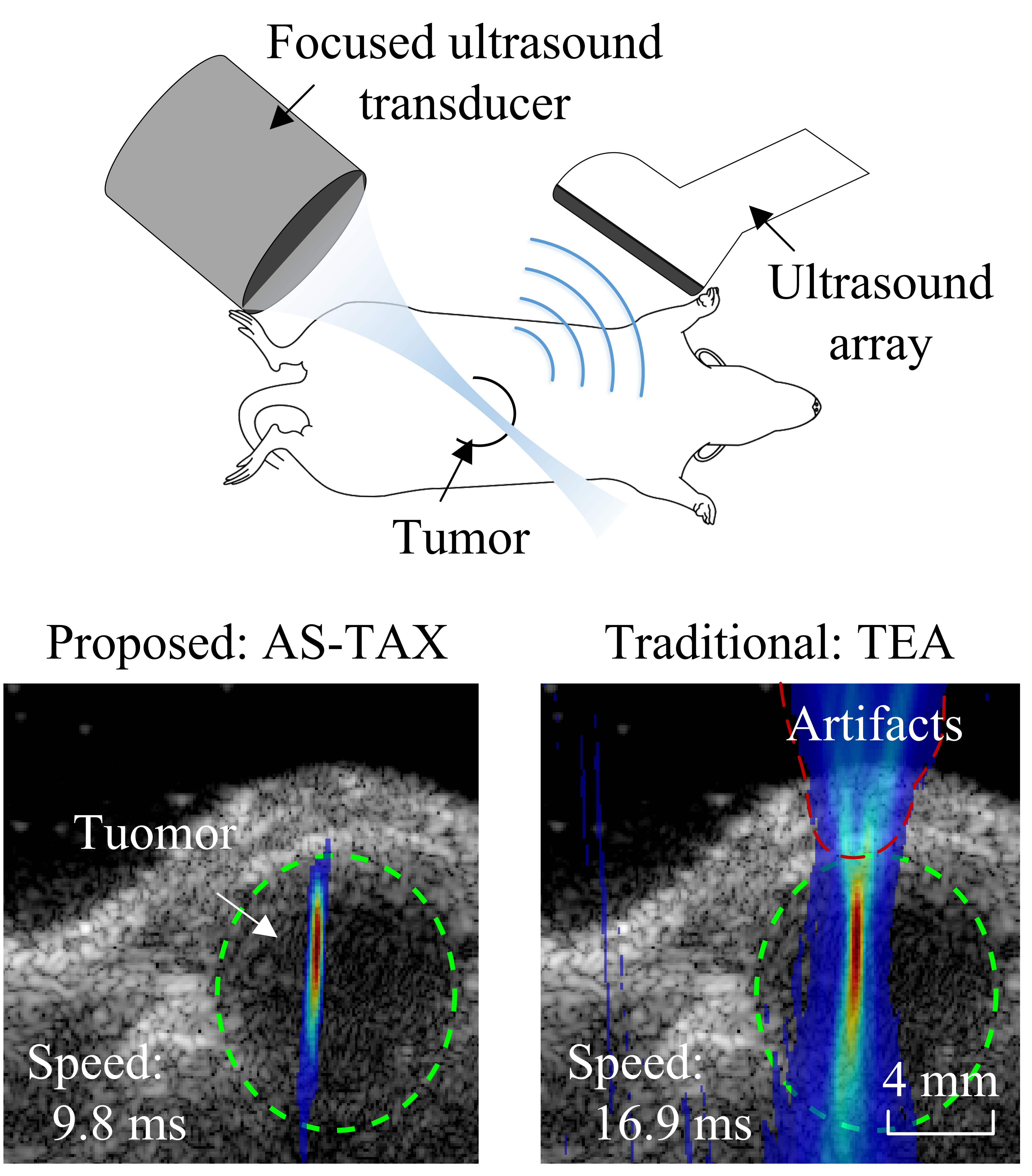}
\end{wrapfigure}%

\begin{abstract}
While passive acoustic mapping (PAM) has been advanced for monitoring acoustic cavitation activity in focused ultrasound (FUS) therapy, achieving both real-time and high-quality imaging capabilities is still challenging.
The angular spectrum (AS) method presents the most efficient algorithm for PAM, but it suffers from artifacts and low resolution due to the diffraction pattern of the imaging array.
Data-adaptive beamformers suppress artifacts well, but their overwhelming computational complexity, more than two orders of magnitude higher than the classical time exposure acoustic (TEA) method, hinders their application in real-time. 
In this work, we introduce the cross-correlated AS method to address the challenge.
This method is based on cross-correlating the AS back-propagated wave fields, in the frequency domain, measured by different apodized sub-apertures of the transducer array to provide the normalized correlation coefficient (NCC) matrix for artifacts suppression.
We observed that the spatial pattern of NCC matrix is variable which can be utilized by the triple apodization with cross-correlation (TAX) with AS scheme, namely the AS-TAX method, for optimal artifacts suppression outcomes.
Both the phantom and mouse tumor experiments showed that: 1) the AS-TAX method has comparable image quality as the data-adaptive beamformers, reducing the energy spread area by 34.8--66.6\% and improving image signal-to-noise ratio by 10.7--14.5 dB compared to TEA; 2) it reduces the computational complexity by two orders of magnitude compared to TEA allowing millisecond-level image reconstruction speed with a parallel implementation; 3) it can well map microbubble cavitation activity of different status (stable or inertial).
The AS-TAX method represents a real-time approach to monitor cavitation-based FUS therapy with high image quality.

\end{abstract}


\bigskip
\end{minipage}}}

\maketitle

\begin{table*}[!t]
\arrayrulecolor{subsectioncolor}
\setlength{\arrayrulewidth}{1pt}
{\sffamily\bfseries\begin{tabular}{lp{6.75in}}\hline
\rowcolor{abstractbg}\multicolumn{2}{l}{\color{subsectioncolor}{\itshape
Highlights}{\Huge\strut}}\\
\rowcolor{abstractbg}$\bullet$ & A cross-correlated AS method combined with triple apodization for high-quality and real-time PAM. \\
\rowcolor{abstractbg}$\bullet${\large\strut} & It suppresses artifacts well and reduces the computational complexity by two orders of magnitude compared to TEA.\\
\rowcolor{abstractbg}$\bullet${\large\strut} & It can well map microbubble cavitation activity of different status with spectral analysis.\\
[2em]\hline
\end{tabular}}
\setlength{\arrayrulewidth}{0.4pt}
\arrayrulecolor{black}
\end{table*}

\section{Introduction}
\label{sec:introduction}

\IEEEPARstart{A}coustic cavitation-based focused ultrasound (FUS) therapy is a valuable non-invasive approach to temporarily open biological barriers, including the blood-brain barrier and blood-tumor barrier, to deliver drugs~\cite{rezai2024ultrasound} and genes~\cite{ilovitsh2020low}, to mechanically destroy tumor tissues~\cite{vidal2022first} and to modulate the brain-macrophage response for immunotherapy~\cite{kline2023characterization}.
To avoid unintended permanent tissue damage associated with cavitation activity and to ensure safe and effective treatment outcomes, the cavitation activity during therapy must be monitored in both space and time.
Passive acoustic mapping (PAM) is a promising technique to address these needs. 

PAM uses the acoustic emissions accompanied with cavitation, which are directly received with an ultrasonic array, to localize and quantify the cavitation activity.
Time exposure acoustics (TEA) is the classical image reconstruction method for PAM~\cite{Norton2000}.
It forms an image based on the "delay and sum" (DAS) operation to estimate the time-varying acoustic wave field in the region-of-interest and calculate the acoustic energy as a representative of the spatial distribution of the cavitation energy. 
The DAS operation was also extended in the frequency domain (FD-TEA)~\cite{haworth2012passive,haworth2016quantitative} to reduce the computational cost, by one order of magnitude, thanks to the reduce number of frequency components compared to the number of temporal samples.
However, as DAS is a pixel-based operation, TEA and FD-TEA are relatively computationally intensive compared to the latter proposed angular spectrum (AS) method~\cite{Arvanitis2017}.
Compared to the DAS-based methods, AS reduces the computational complexity by at least one order of magnitude~\cite{Arvanitis2017}, because it only requires a single propagation step for image reconstruction. 
For PAM with convex arrays, a similar spectral propagator based approach was introduced recently~\cite{Zhu2024PAM}.

Being a frequency domain method like FD-TEA, a particular advantage of AS over TEA in the time domain is that it can selectively use the cavitation-status (stable or inertial) informed frequency components in the signals for image formation, without filtering the signal in advance.
This has been shown particularly useful in developing closed-loop controllers to maintain target levels of stable and inertial cavitation in the space and time \cite{patel2018closed,lee2022}, which may be very important for the clinical translation of acoustic cavitation-based therapies\cite{hu2020focused}.
However, like the TEA method, the image quality of AS also suffers from low spatial resolution and the artifacts, which stem from the diffraction pattern of diagnostic imaging array\cite{haworth2012passive,haworth2016quantitative} and undermine the accuracy of localization and characterization of the cavitation activity. 

To improve the image quality, data-adaptive algorithms for PAM including robust Capon beamformer (RCB)~\cite{Coviello2015}, Eigenspace-based robust Capon beamformer~\cite{Lu2018} and robust beamforming by linear programming~\cite{Lyka2018} have been proposed. These beamformers substantially suppressed artifacts and improved spatial resolution of the image. However, they are highly computationally intensive. 
This is because the process of finding the optimal array steering vector and then the weighting matrix for all the image pixels involves an optimization problem, making the computation cost overwhelming. Therefore, it is still important to explore new approaches offering both fast image reconstruction speed and high image quality for PAM.

Recently, the method dual apodization with cross-correlation (DAX) combined with TEA (TEA-DAX) \cite{Lu2019} was introduced for PAM.
The rationale of DAX is to use a pair of apodization schemes that are highly cross-correlated in the mainlobe but have low or negative cross-correlation in the sidelobe region, which was originally proposed for enhancing the contrast-to-noise ratio in B-mode images~\cite{seo2008sidelobe} and then adapted to be combined with TEA in the time-domain for PAM~\cite{Lu2019}.
TEA-DAX balances well the image quality and computational cost, but is still affected by the X-shape artifacts induced by the weighting matrix associated to the pattern of the apodized aperture.
This issue was latter addressed by combining RCB with DAX~\cite{Lu2020} , which again makes real-time imaging difficult, as RCB is a data-adaptive beamformer.

In this work, we demonstrate that the DAX method can be well combined with frequency domain methods for PAM, e.g. the AS method, for low computational cost.
In the meantime, combining AS with a triple apodization with cross-correlation scheme (AS-TAX) further suppresses the X-shape artifacts.
Both the \emph{in vitro} and \emph{in vivo} experiments showed that the proposed AS-TAX method enables accurate localization of cavitation activities with high image contrast and high temporal resolution.

\section{Methods}
\subsection{Angular spectrum method with DAX}
\label{sec:ASDAX}

\begin{figure}[!t]
\centerline{\includegraphics[width=0.7
\columnwidth]{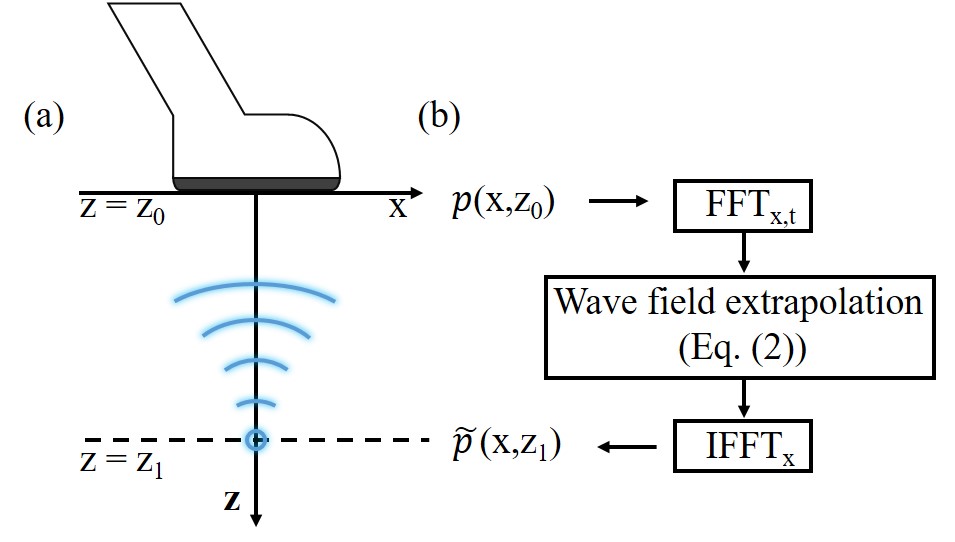}}
\caption{(a) The referential system for PAM. The surface of the ultrasonic array is in the plane $z=z_0$ and its center defines the origin of the coordinate system. The array receives the acoustic emissions from the source located at $z=z_1$. (b) The workflow of the angular spectrum method for PAM.}
\label{fig:ReferenceSystem}
\end{figure}

\begin{figure*}[!t]
\centerline{\includegraphics[width=1.55\columnwidth]{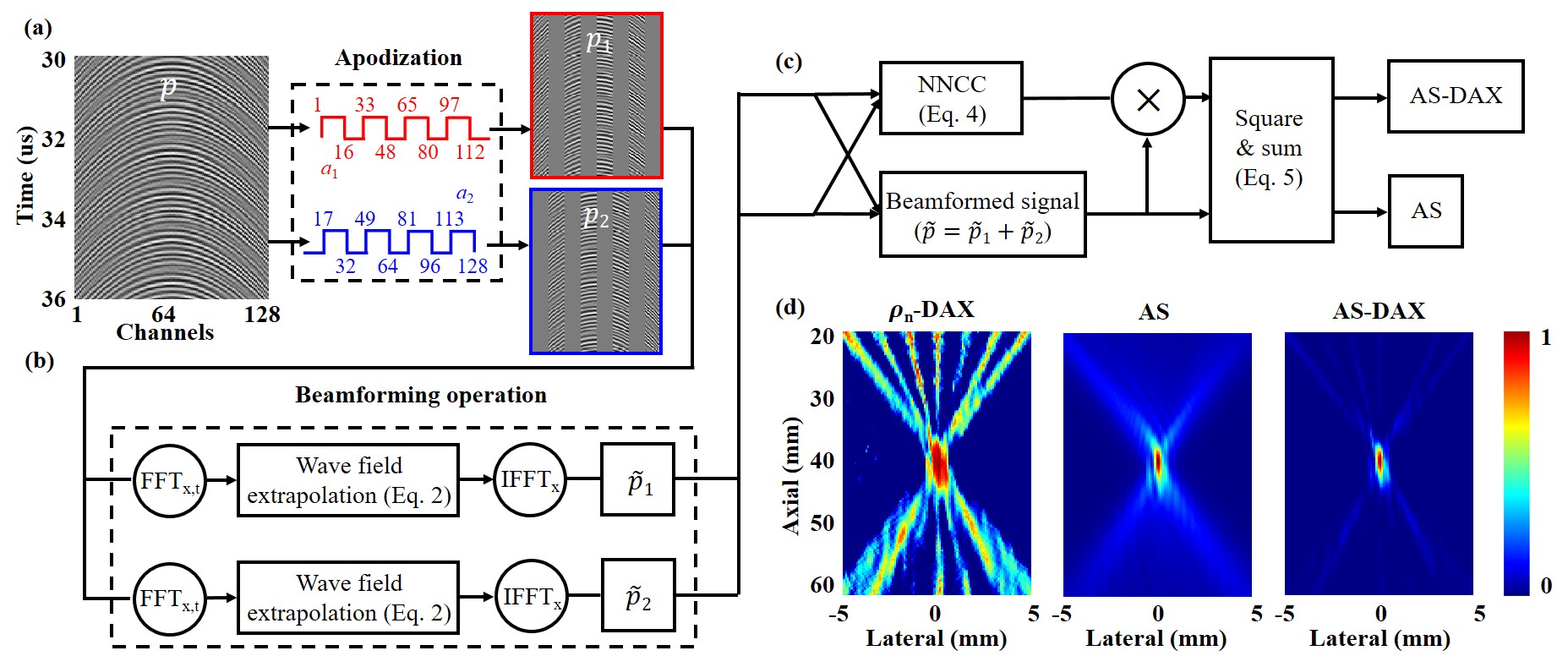}}
\caption{Schematic diagram of the AS-DAX algorithm. (a) The received RF signals ($p$) are firstly weighted by the apodization functions ($a_1$ and $a_2$) yielding two groups of signals ($p_1$ and $p_2$), (b) which are then back-propagated with the AS method to estimate the monochromatic pressure field $\tilde{p}_1$ and $\tilde{p}_2$, respectively. (c) $\tilde{p}_1$ and $\tilde{p}_2$ are then used to calculate the NNCC matrix $\rho_n$, which is combined with the beamformed signals ($\tilde{p}_1+\tilde{p}_2$) for artifacts suppression. (d) Left to right: $\rho_n$, the PAM images formed by the AS and AS-DAX.}
\label{fig:Pipeline}
\end{figure*}

Let $p(\boldsymbol{r},t)$ be the time-varying acoustic pressure at location $\boldsymbol{r}=(x,z)$ in a two-dimensional space, the AS at depth $z$ is denoted as:
\begin{equation}
P(k_x,z,\omega) = \mathcal{F}_x\{\mathcal{F}_t\{p(\boldsymbol{r},t)\}\}
= \mathcal{F}_x\{\tilde{p}(\boldsymbol{r},\omega)\}
\label{FFTAS}
\end{equation}
where $k_x$ is the wave number in the $x$ (lateral) direction, $\omega$ the angular frequency, $\mathcal{F}_x $ and $\mathcal{F}_t $ the Fourier transform over $x$ and $t$, respectively. In a homogeneous medium with no viscosity, the AS at $z=z_1$ can be directly extrapolated by multiplying a spectral propagator with the AS at $z=z_0$ measured by a planar ultrasonic array (Fig.~\ref{fig:ReferenceSystem}) ~\cite{Arvanitis2017}:
\begin{equation}
P(k_x, z_1,\omega) = P(k_x, z_0,\omega)e^{i(z_1-z_0)\sqrt{\omega^2/c^2-k_x^2}}
\label{PropagateAS}
\end{equation}
where $c$ is the sound speed of the medium.
The PAM image $I(\boldsymbol{r})$, with the AS method, is obtained by calculating the energy of the monochromatic pressure field $\tilde{p}(\boldsymbol{r},\omega)$ to represent the spatial distribution of the cavitation energy:
\begin{equation}
I(\boldsymbol{r}) = \sum_{\forall\omega\in\Omega} |\tilde{p}(\boldsymbol{r},\omega)|^2 
\label{PAMAS}
\end{equation}
where $\Omega$ defines the selected frequencies. Thanks to the extensive use of fast Fourier transform (FFT), the computational cost of the AS method is reduced by at least one order of magnitude compared to the TEA~\cite{Arvanitis2017}. However, the image quality is not improved.
Here, we extend the DAX technology~\cite{Lu2019,Lu2020} for PAM to the frequency domain to suppress artifacts and improve image quality (Fig.~\ref{fig:Pipeline}). 

The received radio frequency (RF) signals $p(x,z_0,t)$ are firstly weighted by two complementary apodization functions ($\boldsymbol{a}_1$ and $\boldsymbol{a}_2$) along the direction of transducer elements, which become two groups of signals, $p_1(x,z_0,t)$ and $p_2(x,z_0,t)$. In $\boldsymbol{a}_1$ and $\boldsymbol{a}_2$, each $m$ adjacent active elements are interleaved with $(n-1)m$ inactive elements, where $n=2$ is the number of the apodization functions. Correspondingly, the pattern of the selection is defined as Pattern-m in this work.  For instance, for an array of 8 elements with Pattern-2, the complementary apodization functions ($n=2$) are $\boldsymbol{a}_1=[1, 1, 0, 0, 1, 1, 0, 0]$ and $\boldsymbol{a}_2=[0, 0, 1, 1, 0, 0, 1, 1]$.
The weighted RF signals are then used to propagate $\tilde{p}(\boldsymbol{r},\omega)$ for every image pixel using the AS method (Eq.~\eqref{PropagateAS}). Then, the normalized cross-correlation coefficient (NCC) between the two monochromatic pressure fields $\tilde{p}_1(\boldsymbol{r},\omega)$ and $\tilde{p}_2(\boldsymbol{r},\omega)$ is calculated for every pixel:


\begin{equation}
\begin{aligned}
\rho (\boldsymbol{r})=  \frac{\tilde{p}^{H}_1 \tilde{p}_2}{\sqrt{\tilde{p}^{H}_1\tilde{p}_1}\sqrt{\tilde{p}^{H}_2\tilde{p}_2}}
\label{CorrelationCoeffAS}
\end{aligned}
\end{equation}
Where $(.)^H$ is the conjugate transpose operation, and $\tilde{p}$ represents $\tilde{p}(\boldsymbol{r},\omega)$ for simplicity. When $\rho (\boldsymbol{r})$ is smaller than the threshold value $\varepsilon$ (usually 0.001), it is forced to equal $\varepsilon$ to avoid very small value~\cite{Lu2020,Lu2019}.
In previously proposed DAX-based methods \cite{Lu2020, Lu2019, seo2008sidelobe}, cross-correlating two wave fields (Eq.~\eqref{CorrelationCoeffAS}) were carried out in the temporal domain. In Appendix\ref{sec:CoefficientDerivation}, we show that $\rho(\boldsymbol{r})$ calculated in temporal and frequency domain are equivalent for received signals with 0 offsets.
Finally, the maximum value of $\rho(\boldsymbol{r})$ is normalized to 1 which becomes normalized NCC (NNCC) $\rho_n(\boldsymbol{r})$ and the PAM image is formed by:
\begin{equation}
\begin{aligned}
I(\boldsymbol{r})= \sum_{\forall\omega\in\Omega}{(\rho_{n}(\boldsymbol{r})\vert\tilde{p}_1(\boldsymbol{r},\omega)+\tilde{p}_2(\boldsymbol{r},\omega)\vert)}^2
\label{EnergyASDAX}
\end{aligned}
\end{equation}
This method (Eq.~(\ref{EnergyASDAX})) is termed the AS-DAX method.

\subsection{Time exposure acoustics with DAX}
\label{sec:TEADAX}
TEA \cite{Norton2000} reconstructs PAM images based on the DAS operation. The delayed element signals are firstly summed across the receiving channels coherently, as a representation of the time-varying wave field at location $\boldsymbol{r}$:
\begin{equation}
p(\boldsymbol{r},t)=\sum_{l=1}^{L}d_l(\boldsymbol{r})s_l(t+\tau_l(\boldsymbol{r}))
\label{DelayTEA}
\end{equation}
where $L$ is the number of elements, $l$ indexes the transducer elements, $s_l(t)$ is the passively received RF signal by the $l$th element, $d_l(\boldsymbol{r})$ and $\tau_l(\boldsymbol{r})$ are the traveling distance and time of the sound wave from location $\boldsymbol{r}$ to the $l$th element, respectively.
Then, the energy of $p(\boldsymbol{r},t)$ over an interested time interval [$T_0$,$T_0+\triangle T$] is calculated to represent the cavitation source energy at location $\boldsymbol{r}$:
\begin{equation}
I(\boldsymbol{r})=\sum_{t=T_0}^{T_0+\triangle T} p(\boldsymbol{r}, t)^2
\label{PAMTEA}
\end{equation}

To apply the DAX technique\cite{Lu2019}, the two wave fields $p_1(\boldsymbol{r}, t)$ and $p_2(\boldsymbol{r}, t)$ obtained with the RF signals weighted by $\boldsymbol{a}_1$ and $\boldsymbol{a}_2$ are cross-correlated to calculate the NCC:

\begin{equation}
\begin{aligned}
\rho (\boldsymbol{r})=  
\frac{(p_1-\overline{p}_1)^{T} (p_2-\overline{p}_2)}{\sqrt{(p_1-\overline{p}_1)^{T}(p_1-\overline{p}_1)}\sqrt{(p_2-\overline{p}_2)^{T}(p_2-\overline{p}_2)}}
\label{CorrelationCoeff}
\end{aligned}
\end{equation}
Where $(.)^T$ is the transpose operation, $\overline{(.)}$ is the mean operation, $p$ represents $p(\boldsymbol{r},t)$ for simplicity.
As in the AS-DAX method, the normalized $\rho_n(\boldsymbol{r})$ is finally applied to the TEA formed images:
\begin{equation}
\begin{aligned}
I(\boldsymbol{r})=\sum_{t=T_0}^{T_0+\triangle T} (\rho_n(\boldsymbol{r})(p_1(\boldsymbol{r},t)+p_2(\boldsymbol{r},t)))^2
\label{PAMDAXTEA}
\end{aligned}
\end{equation}
This method (Eq.~\eqref{PAMDAXTEA}) is named the TEA-DAX method.

\begin{figure*}[!t]
\centerline{\includegraphics[width=1.55\columnwidth]{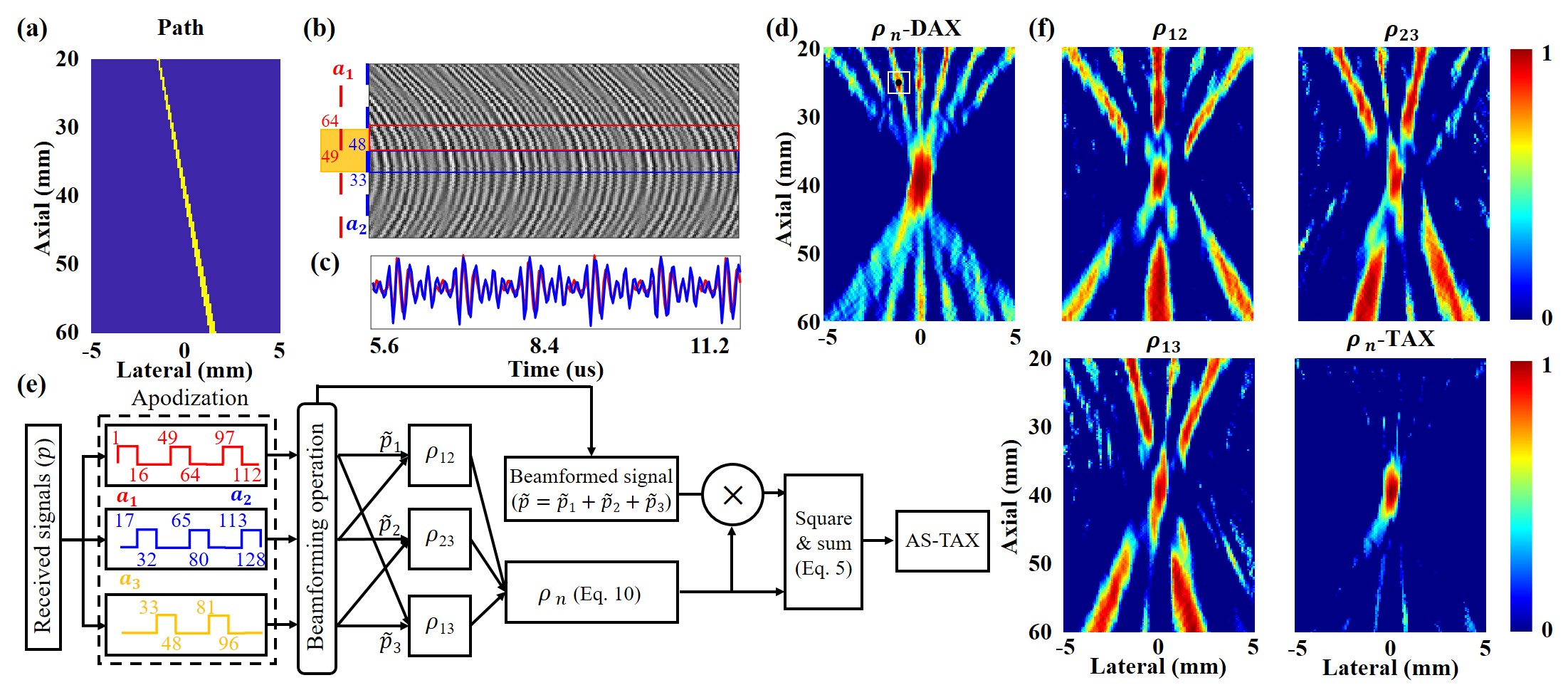}}
\caption{(a) A path along which the delayed signals for channel 33--64 are highly correlated. (b) The delayed signals for a pixel located at (-1.1 mm, 26.2 mm) on the path delineated in (a). (c) The beamforming signals obatined using the apodization functions $\boldsymbol{a}_1$ (red) and $\boldsymbol{a}_2$ (blue). (d) An example of the NNCC matrix obtained with the DAX operation. (e) The workflow the AS-TAX algorithm. (f) The three NCC matrix $\rho_{12}$, $\rho_{23}$, $\rho_{13}$ and the combined NNCC matrix used in TAX.}
\label{fig:DAXTAXCoef}
\end{figure*}

\subsection{Angular spectrum method with TAX}
\label{sec:TripleApod}
As noted previously\cite{haworth2016quantitative,Lu2020}, the DAX technology for PAM is challenged by the X-shape artifacts (Fig.~\ref{fig:DAXTAXCoef}(d)). 
This is mainly because the delayed signals for a few channels (see an example in Fig.~\ref{fig:DAXTAXCoef}(b)) have similar phase along the path of the X-shape artifacts (Fig.~\ref{fig:DAXTAXCoef}(a)).
Thus, when these delayed signals are assigned to different apodized sub-arrays using apodization functions, the beamformed signals for these sub-arrays are highly correlated (Fig.~\ref{fig:DAXTAXCoef}(c)) and exhibit a high NCC value along the path, which contributes to the X-shape artifacts (Fig.~\ref{fig:DAXTAXCoef}(d)). This issue was overcame by using the RCB beamformer for calculating a pre-weighted version of $ p(\boldsymbol{r},t)$ at the cost of high computational complexity \cite{Lu2020}. In this work, we propose to use the triple apodization cross-correlation (TAX) approach, i.e., the transducer aperture is divided into three sub-arrays, to address this issue. This approach only adds a small amount of computational cost while the X-shape artifacts can be well suppressed.

In TAX, the received signals are weighted by three complementary apodization functions ($\boldsymbol{a}_1, \boldsymbol{a}_2, \boldsymbol{a}_3$) for beamforming (Eq.~\eqref{PropagateAS}). For an array of 12 elements with Pattern-2 ($m=2$), the apodization functions ($n=3$) are $\boldsymbol{a}_1$ = [1, 1, 0, 0, 0, 0, 1, 1, 0, 0, 0, 0], $\boldsymbol{a}_2 = $[0, 0, 1, 1, 0, 0, 0, 0, 1, 1, 0, 0] and $\boldsymbol{a}_3 = $[0, 0, 0, 0, 1, 1, 0, 0, 0, 0, 1, 1]. The beamformed signals ($\tilde{p}_1(\boldsymbol{r},\omega)$, $\tilde{p}_2(\boldsymbol{r},\omega)$ and $\tilde{p}_3(\boldsymbol{r},\omega)$) are paired to calculate the NCCs: $\rho_{12}$ for the pair of $(\tilde{p}_1(\boldsymbol{r},\omega)$ and $\tilde{p}_2(\boldsymbol{r},\omega)$, $\rho_{23}$ for $\tilde{p}_2(\boldsymbol{r},\omega)$ and $\tilde{p}_3(\boldsymbol{r},\omega)$, and $\rho_{31}$ likewise, with Eq.~\eqref{CorrelationCoeffAS} (Fig.~\ref{fig:DAXTAXCoef}(e)). As these beamformed signals are highly correlated near the cavitation sources while the spatial pattern of the X-shape artifacts in the NCC matrix varies between different pairs (Fig.~\ref{fig:DAXTAXCoef}(f)), this artifact is well suppressed by multiplying $\rho_{12}$, $\rho_{23}$ and $\rho_{31}$ together. Thus, we used the following equation to calculate the weights for TAX:
\begin{equation}
\begin{aligned}
\rho_{TAX}(\boldsymbol{r})=\sqrt[3]{(\rho_{12}(\boldsymbol{r})\rho_{23}(\boldsymbol{r})\rho_{31}(\boldsymbol{r}))}
\label{CorrelationCoeffTAX}
\end{aligned}
\end{equation} 
Then the coefficients $\rho_{TAX}$ are also normalized to be $\rho_n$ and the PAM image is formed by:
\begin{equation}
\begin{aligned}
I(\boldsymbol{r})= \sum_{\forall\omega\in\Omega}{(\rho_{n}(\boldsymbol{r})\vert\tilde{p}_1(\boldsymbol{r},\omega)+\tilde{p}_2(\boldsymbol{r},\omega)+\tilde{p}_3(\boldsymbol{r},\omega)\vert)}^2
\label{EnergyASTAX}
\end{aligned}
\end{equation}
This method (Eq.~(\ref{EnergyASTAX}) is termed the AS-TAX method.

\subsection{Experiments}
\subsubsection{Experimental Setup}
\label{sec:ExpSetup}
\begin{figure}[!t]
\centerline{\includegraphics[width=0.95\columnwidth]{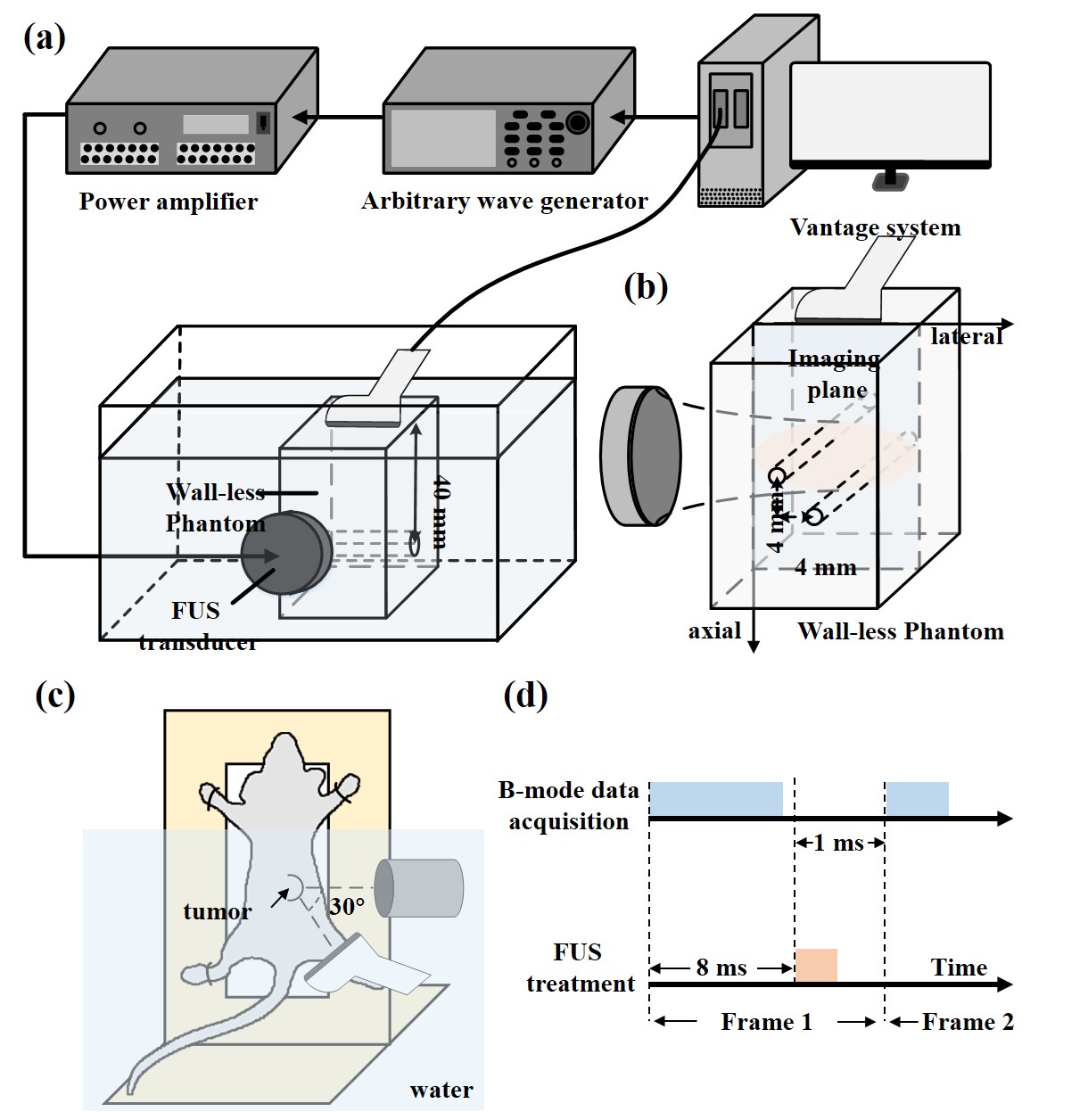}}
\caption{The experimental setup for PAM using (a) a wall-less phantom, (b) flow tube and (c) mice experiment. (d) The sequence for FUS treatment and data acquisitions for B-mode imaging and PAM.}
\label{fig:Setup} 
\end{figure}

All the experiments were carried out in degassed and deionized water. A 128-element linear transducer array (CL15-7, Philips, Netherlands; pitch: 0.178 mm) was employed for B-mode imaging and PAM in the experiments. The array was interfaced to a research ultrasound system (Vantage 256, Verasonics, USA) to record the RF signals and synchronize the FUS experiments. To initiate acoustic emissions from the MBs (Vevo MicroMarker, Visualsonics, Canada), two FUS transducers of different center frequencies (Guangzhou Doppler Electronic Technologies, China) were driven by the power amplifier (ATA-1372A, Aigtek, China) with a pulsed length of 100 cycles. The peak-negative-pressure (PNP) at the focus of the FUS transducer measured by a needle hydrophone (NH0500, Precision Acoustics, U.K.) was between 0.1--2 MPa in deionized and degassed water. The sequences for B-mode imaging (11 plane waves between -20$^\circ$--20$^\circ$) and FUS treatment were interleaved (Fig.~\ref{fig:Setup} (d)). For each acquisition, the cavitation signals (2048 times samples) were sampled at 35.6 MHz.

\subsubsection{Phantom Experiments}
\label{sec:PhantomExp}

Wall-less phantoms (2\% agar w/v + 0.2\% cellulose powder w/v) with a tunnel (Phantom 1) and two tunnels (Phantom 2) were made to evaluate the proposed method. In Phantom 1, a tunnel (5 mm diameter) was positioned 40 mm below the surface of the phantom where the imaging array was placed parallel to the tunnel (Fig.~\ref{fig:Setup} (a)). In Phantom 2, two parallel tunnels (2 mm diameter), separated by 8 mm both laterally and axially, were placed 34 and 42 mm below the surface of the phantom and the imaging array was positioned perpendicularly to the tunnels to image their cross-sections (Fig.~\ref{fig:Setup} (b)). The MBs flowed (1$\times10^7$ MBs/ml) in the tunnels at a constant speed (10 ml/min) pushed by a syringe pump (LSP01-1B, Baoding Ditron Electronic Technology Co., Ltd., China). To initiate cavitation activity in Phantom 1, the 1 MHz FUS transducer (focal depth: 24 mm, aperture diameter: 24 mm) was used, and the PNP was 750 kPa. In Phantom 2, to initiate cavitation activities in both tunnels simultaneously, the 0.5 MHz FUS transducer (focal depth: 64 mm, aperture diameter: 64 mm) with a larger focal zone (-3 dB beam width (21.9 mm, 3.1 mm)), was tilted at an angle to the horizontal plane and the focus was place in the middle of the two tunnels.
To warrant excitation of multiple cavitation sources for imaging, the PNP was increased to 1.5 MPa because the two tunnels were not at the focus thus their experienced PNPs were lower. We also varied the acoustic pressure (PNP: 0.1-2 MPa) in Phantom 1 to initiate acoustic emissions of different cavitation status, i.e. moderate cavitation, stable cavitation and inertial cavitation, to demonstrate that, based on spectral analysis for PAM~\cite{vignon2013,Zhu2024PAM}, the AS-TAX method can precisely localize different types of cavitation activity in the space and time (Sec~\ref{sec:ArtifactsEval}).

\subsubsection{Mouse Experiments}
\label{sec:MouseExp}

Hepa1-6 liver cancer cells (10$^7$ cells, 50 $\mu$L) were subcutaneously injected into two C57BL/6 mice (male, 5--6 weeks old, weighing 20--22g), and allowed to form tumors sized 200--400 mm$^3$. During FUS treatments (PNP: 750 kPa), the mice, anesthetized by tribromoethanol (30 $\mu$l/g), were tied to a shelf and placed in water. The imaging array was facing the tumor on which the 1 MHz FUS transducer and the imaging array were co-registered. A bolus of 50 $\mu$L MB solution (Sonazoid, GE Healthcare, Norway) was injected into the tumor. All animal experiments were approved by the Institutional Animal Care and Use Committee of the ShanghaiTech University (Protocol 20220516001).

\subsection{Image Reconstruction and Evaluation Metrics}
\label{sec:ImageRecon}

All the images were reconstructed with a pixel size of ($\Delta_x$ = 0.089 mm, $\Delta_z$ = 0.2 mm) in Matlab (R2022A, MathWorks, Inc., USA) on a workstation (Intel Xeon 3 GHz, 128 GB memory) interfaced with a graphical processing unit (GPU; NVIDIA GeForce RTX 4090). For the AS-based methods, the received RF signals were firstly interpolated by applying 2 times zero-stuffing in the channel direction, and then were apodized for beamforming using the DAX or TAX technology. We also implemented the CUDA C++ program for all the PAM reconstruction methods for parallelization using the GPU. In the code, all the independent and parallelizable calculation, including DAS (Eq.~\eqref{DelayTEA}), calculation of NCC (Eq.~\eqref{CorrelationCoeff}) and estimation of energy by summing over $t$ (Eq.~\eqref{PAMDAXTEA}) in TEA-DAX, and extrapolation of AS (Eq.~\eqref{PropagateAS}), calculation of NCC (Eq.~\eqref{CorrelationCoeffAS}) and estimation of energy by summing over $\omega$ (Eq.~\eqref{EnergyASDAX}) in the AS-DAX and AS-TAX method were assigned to a thread on the GPU. 

To compare with the (time domain) TEA-based methods, wideband frequency components (3--15 MHz, 690 frequency bins) covering the working bandwidth of the CL15-7 were used for the AS-based methods. To test whether AS-DAX and AS-TAX could suppress artifacts when only the cavitation status informed frequency components are used, i.e. whether they can be used to localize cavitation activity of different status, we also reconstructed the images with only the harmonic (each centered at 4.0, 5.0,..., 13.0 and spans ± 0.1 MHz), ultra-harmonic (each centered at 4.5,..., 13.5 and spans ± 0.1 MHz) or broadband (each centered at 3.75, 4.25,..., 13.25 and spans ± 0.05 MHz) frequency components of the cavitation signals excited at 0.1 MPa, 0.8 MPa and 2 MPa (PNP) in Phantom 1, respectively. The number of frequency bins used for the harmonic, ultra-harmonic, and broadband components was 128 each.

We compared the AS-DAX, AS-TAX and TEA-based methods in terms of energy spread area (ESA), artifacts suppression capability, localization accuracy and computational complexity. The ESA was evaluated as the $A_{0.5}$ area where the pixel values were greater than half of the maximum value in the image \cite{abadi2018frequency,Lu2019,Lu2020}. Artifacts suppression capability was quantified as the image signal-to-noise ratio (ISNR):
\begin{equation}
\label{EQ_SNR}
ISNR = 10\log_{10}(I_{\mathrm{Inside}}/I_{\mathrm{Outside}})
\end{equation}
where $I_{\mathrm{Inside}}$ and $I_{\mathrm{Outside}}$ are the mean pixel value inside and outside $A_{0.5}$ area. 

To evaluate source localization accuracy, the centroid of the cavitation energy in the PAM images is defined as:
\begin{equation}
\boldsymbol{c}_{I} = \frac{\sum_{i=1}^N I_i \boldsymbol{r}_i}{\sum_{i=1}^N I_i}, \quad I_i > \alpha\cdot\operatorname{max}(I(\boldsymbol{r}))
\label{Centroid}
\end{equation}
where $\alpha=0.7$ is the threshold to avoid the influence of the artifacts on source localization. 
We also find the centroid $\boldsymbol{c}_{\rho}$ in the map of NNCC matrix with the assumption that the centroid should be in a region of very high NNCC value ($>0.9$), to help the determination of the apodization functions.
In this case, the $I$ in Eq.~\ref{Centroid} was replaced with $\rho$ and $\alpha=0.9$.

For computational complexity, the time and space complexity and the run time of the methods to reconstruct an image (76800 pixels) on the CPU using a single thread (without parallelization) and GPU were evaluated for comparison. 
We noticed that the spectral propagator in the AS-based method and the delays used in the TEA-based methods do not change for fixed imaging settings which can be pre-calculated and stored in look-up tables prior to image reconstruction. Therefore, we separately evaluated the pre-run time for calculating the spectral propagator or the delays, and the run time for image reconstruction in the comparison.

For statistical analysis, the evaluation metrics were computed on 100 measurements for each experimental setting. The run time of the methods was evaluated 10 times on the CPU and 100 times on the GPU.

\section{Results}

\subsection{Selection of the apodization functions}
\label{sec:SelectionApod}
\begin{figure}[!t]
\centerline{\includegraphics[width=0.9\columnwidth]{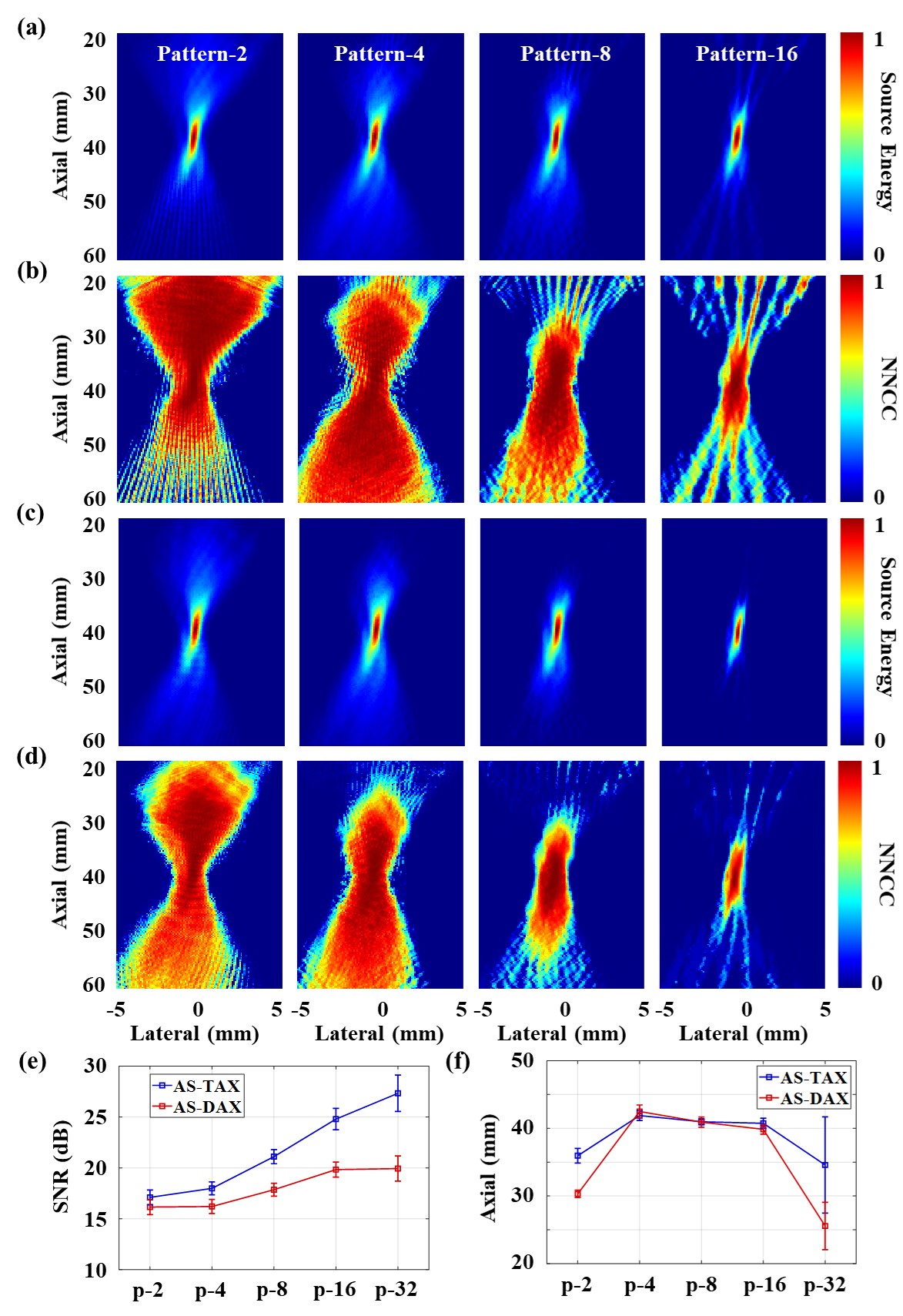}}
\caption{PAM images reconstructed by (a) AS-DAX and (c) AS-TAX with different patterns using the experimental data from Phantom 1. The NNCC matrices of (b) AS-DAX and (d) AS-TAX used associated with the weighted images in (a) and (c), respectively. (e) ISNR and (f) axial location of the centroid of the NNCC matrices calculated in AS-DAX and AS-TAX. On the horizontal axis, p-$m$ denotes Pattern-$m$.}
\label{fig:DifferentAperture}
\end{figure}

Using the experimental data from Phantom 1, we determined the suitable complementary apodization functions for our experimental settings to localize cavitation source with the cross-correlation based methods.
The results (evaluated on 100 measurements with wideband frequency components) showed that the artifacts suppression capability of the cross-correlation based methods for PAM is closely related to the apodization functions (Fig.~\ref{fig:DifferentAperture}). 

For AS-DAX, the NNCCs in the artifact region were mostly above 0.9 with Pattern-2, Pattern-4 and Pattern-8 (Fig.~\ref{fig:DifferentAperture}(b)), resulting in weak artifact suppression capability (Fig.~\ref{fig:DifferentAperture}(a)).
With Pattern-16, the area of high NNCC value ($\ge$ 0.9) was better confined to the location of the cavitation source (Fig.~\ref{fig:DifferentAperture}(b)).
Thus, the artifacts were effectively suppressed in the PAM images (Fig.~\ref{fig:DifferentAperture}(a)). 
These observations are consistent with the evolution of the values in the ISNR and the estimated source location (centroid) in the NNCC matrices (Fig.~\ref{fig:DifferentAperture}(e-f)).
While further increasing the $m$ value to 32 in the apodization functions improved the ISNR (Fig.~\ref{fig:DifferentAperture}(e)), the estimated source location deviated from the reference location (depth at $z=40$ mm) by about 15 mm in average (Fig.~\ref{fig:DifferentAperture}(f)).
This was because the NNCC values in the X-type artifacts region became excessively high which were higher than those near the cavitation sources.
For AS-TAX, the aforementioned observations for AS-DAX also apply and the former method showed better artifacts suppression capability compared to the latter, with the optimum performance tied to Pattern-16 (Fig.~\ref{fig:DifferentAperture}(c-d)).
Consequently, we selected Pattern-16 for the cross-correlation based methods in this work.

\subsection{Artifacts suppression evaluation}
\label{sec:ArtifactsEval}
As presented in Appendix\ref{sec:FDTEA}, DAX and TAX schemes can be combined with TEA, frequency domain (FD) TEA and AS for PAM and the image quality of each combination and their counterparts are equivalent.
We, therefore, focus on comparing TEA, TEA-DAX, AS-DAX and AS-TAX to emphasize on the artifacts suppression capability leveraged by the DAX and TAX schemes in the following.

\begin{figure}
\centerline{\includegraphics[width=1.0\columnwidth]{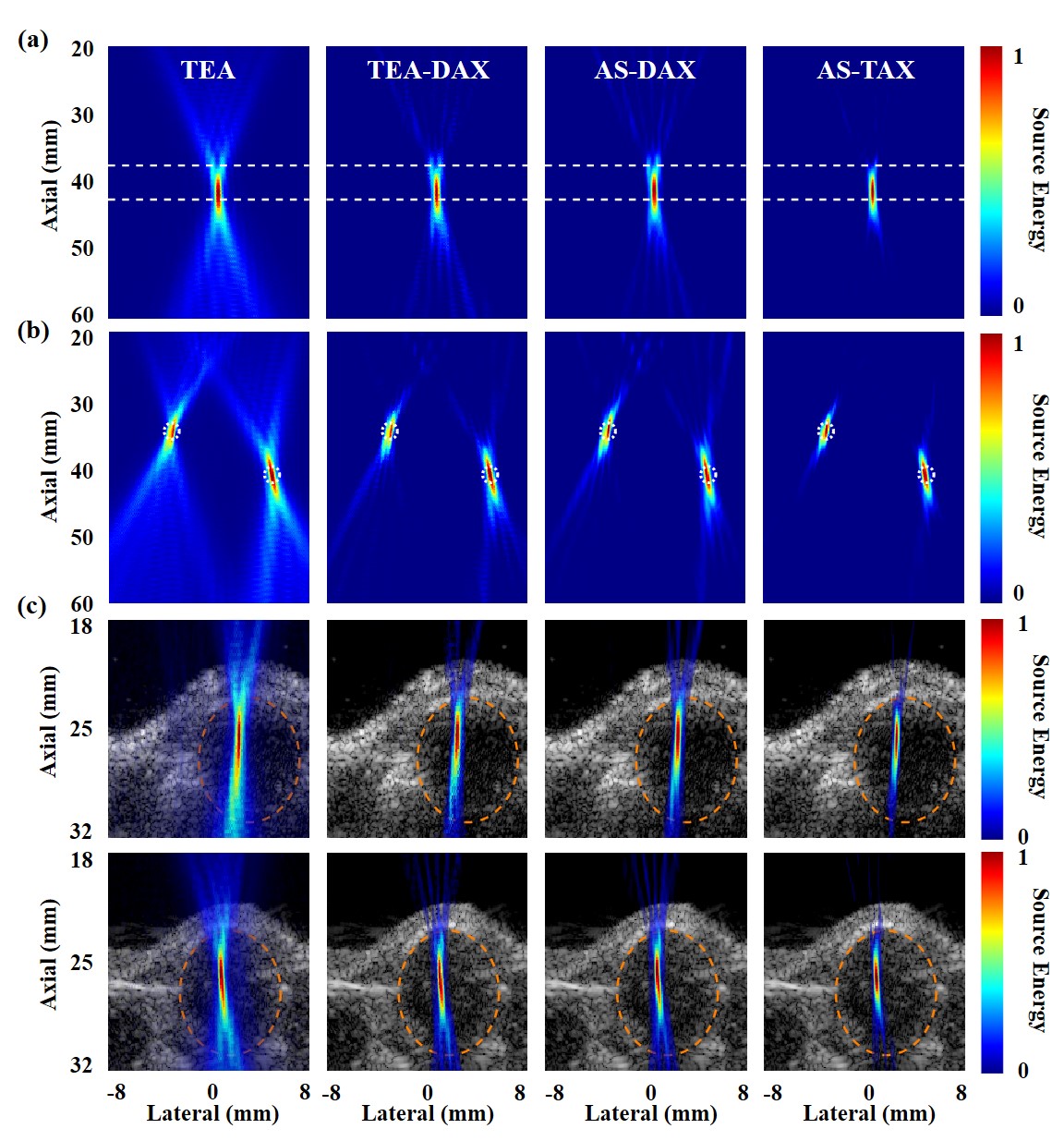}}
\caption{PAM images reconstructed by the TEA, TEA-DAX, AS-DAX and AS-TAX methods in (a) Phantom 1, (b) Phantom 2, and (c) \emph{in vivo} mouse tumors (from two mice). The white dash lines and circles in (a) and (b) outline the tunnel's boundary in the longitudinal and transverse directions, respectively. The orange dash circle in (c) outlines the tumor's boundary. The dynamic range of the B-mode images is 60 dB.}
\label{fig:ExperimentalResultsDifferentMethod}
\end{figure}

The comparison of the PAM images reconstructed by the TEA, TEA-DAX, AS-DAX and AS-TAX methods using widedband frequencies (690 frequency bins) showed that the artifacts associated with the TEA method can be well suppressed (Fig.~\ref{fig:ExperimentalResultsDifferentMethod}, Table~\ref{table:SimulatedAndExperimentalPerformance}), with the optimum results provided by the TAX-based method in our data. 
In Phantom 1, for imaging single cavitation source, the average ESA ($A_{0.5}$) and the ISNR were reduced from 5.94 $mm^2$ (with TEA) to 2.14 $mm^2$ (with AS-TAX) and improved from 12.23 dB (with TEA) to 22.89 dB (with AS-TAX), respectively (Table~\ref{table:SimulatedAndExperimentalPerformance}), as the AS-TAX method better preserved most cavitation energy within a small area.
The difference in the images reconstructed by the TEA-DAX and AS-DAX methods was marginal (Fig.~\ref{fig:ExperimentalResultsDifferentMethod}), which is consistent with the observed small differences in the average value of ESA and ISNR of the two images, and the differences were 0.29 $mm^2$ and 0.46 dB (or 7.7$\%$ and 2.5$\%$ relative to the values of TEA-DAX), respectively (Table~\ref{table:SimulatedAndExperimentalPerformance}).
The average centroid of the cavitation energy in the PAM images reconstructed by TEA, TEA-DAX, AS-DAX and AS-TAX showed negligible differences, less than 0.11 mm and 0.41 mm in the lateral and axial directions, which were (-0.48 mm, 40.98 mm), (-0.50 mm, 40.82 mm), (-0.58 mm, 40.57 mm) and (-0.59 mm, 40.61 mm), respectively.
We also compared the images reconstructed by AS-TAX with the ones by the data-adaptive beamformers, including DAX-RCB, EISRCB and RLPB (Appendix\ref{sec:ComparisionAdaptive}).
The results showed that their differences in the localized cavitation source location (centroid) and ISNR were less than 0.32 mm and 3.74 dB, indicating their comparable source localization and artifacts suppression capability.

In Phantom 2, for imaging multiple cavitation sources, the observations are consistent with the ones in Phantom 1.
The average ESA was reduced from 9.02 $mm^2$ (with TEA) to 3.01 $mm^2$ (with AS-TAX), and the ISNR was improved from 6.97 dB (with TEA) to 21.46 dB (with AS-TAX), respectively (Table~\ref{table:SimulatedAndExperimentalPerformance}).
The marginal difference in the images of TEA-DAX and AS-DAX (Fig.~\ref{fig:ExperimentalResultsDifferentMethod}) was also evidenced by the average ESA and ISNR of the two methods (Table~\ref{table:SimulatedAndExperimentalPerformance}).
The remained X-shape artifacts in the DAX-based methods were suppressed by the AS-TAX method (Fig.~\ref{fig:ExperimentalResultsDifferentMethod}(b)).
The two cavitation sources in the images reconstructed by TEA, TEA-DAX, AS-DAX and AS-TAX were localized at  (-3.12 mm, 34.91 mm) vs (5.37 mm, 41.93 mm), (-3.12 mm, 34.91 mm) vs (5.35 mm, 41.74 mm), (-3.20 mm, 34.80 mm) vs (5.22 mm, 41.42 mm) and (-3.21 mm, 34.84 mm) vs (5.23 mm, 41.50 mm), respectively.
Thus, the differences were small which were less than 0.15 mm and 0.51 mm in the lateral and axial directions, respectively.
Note that in Phantom 2, the images were cropped at $x=0$ mm to separate the two cavitation sources and the centroid was estimated in each cropped image.

\begin{table}
\centering
\caption{ESA and ISNR (mean $\pm$ standard deviation) of the TEA, TEA-DAX, AS-DAX and AS-TAX methods using the \emph{in vitro} Phantom and \emph{in vivo} mouse experimental data.}
\label{table:SimulatedAndExperimentalPerformance}
\setlength{\tabcolsep}{5pt}
\begin{tabular}{c|c|c|c}
\hline\hline
Data Type & Algorithm & {\makecell[c]{ESA ($mm^2$)}} & {\makecell[c]{ISNR (dB)}} \\
\hline
\multirow{4}{*}{\makecell[c]{Phantom 1 \\ (single source)}} & TEA
& 5.94 $\pm$ 2.56 & 12.23 $\pm$ 0.79 \\
\multirow{5}{*}{} & TEA-DAX
& 3.77 $\pm$ 1.34 & 18.50 $\pm$ 0.75 \\
\multirow{5}{*}{} & AS-DAX
& 3.48 $\pm$ 1.35 & 18.96 $\pm$ 0.75 \\
\multirow{5}{*}{} & AS-TAX
& 2.14 $\pm$ 0.76 & 22.89 $\pm$ 1.06 \\
\hline
\multirow{4}{*}{\makecell[c]{Phantom 2 \\ (multiple sources)}} & TEA
& 9.02 $\pm$ 3.85 & 6.97 $\pm$ 1.38 \\
\multirow{5}{*}{} & TEA-DAX
& 3.91 $\pm$ 1.49 & 18.25 $\pm$ 0.65 \\
\multirow{5}{*}{} & AS-DAX
& 4.46 $\pm$ 1.50 & 18.14 $\pm$ 0.51 \\
\multirow{5}{*}{} & AS-TAX
& 3.01 $\pm$ 1.10 & 21.46 $\pm$ 0.66 \\
\hline
\multirow{4}{*}{\makecell[c]{\emph{in vivo} \\ mouse tumor}} & TEA
& 0.89 $\pm$ 0.92 & 13.74 $\pm$ 0.72 \\
\multirow{5}{*}{} & TEA-DAX
& 0.66 $\pm$ 0.70 & 21.22 $\pm$ 0.83 \\
\multirow{5}{*}{} & AS-DAX
& 0.69 $\pm$ 0.72 & 21.08 $\pm$ 0.85 \\
\multirow{5}{*}{} & AS-TAX
& 0.58 $\pm$ 0.61 & 26.68 $\pm$ 1.03 \\
\hline\hline
\end{tabular}
\end{table}

We further compared the methods using \emph{in vivo} mouse tumor data to validate our observations (Fig.~\ref{fig:ExperimentalResultsDifferentMethod}(c)).
Consistently, compared to TEA, the average ESA was reduced from 0.89 $mm^2$ to 0.58 $mm^2$ and ISNR was improved from 13.74 dB to 26.68 dB by AS-TAX, respectively. Consistently, the difference between the TEA-DAX and AS-DAX methods was marginal (Table~\ref{table:SimulatedAndExperimentalPerformance}). 

\begin{figure*}
\centerline{\includegraphics[width=2.0\columnwidth]{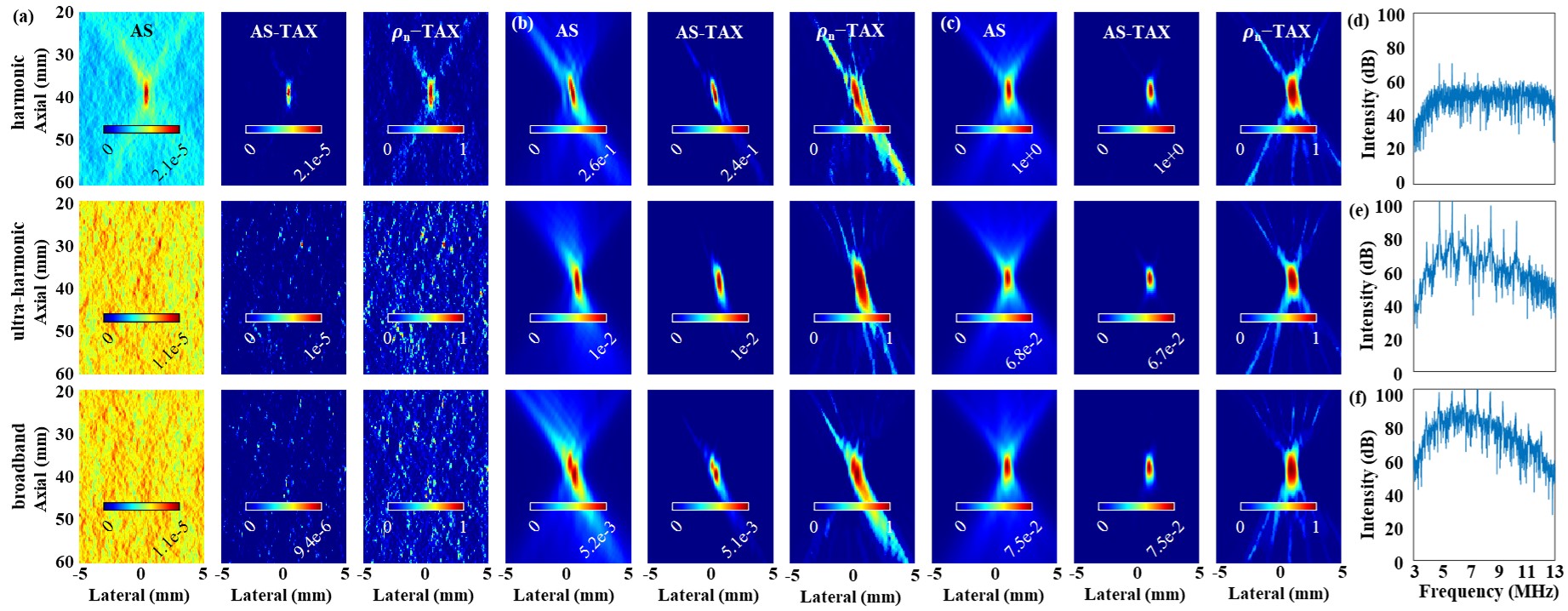}}
\caption{PAM images reconstructed by AS and AS-TAX, and the NNCC matrices of AS-TAX ($\rho_n$-TAX) with harmonic, ultra-harmonic and broadband frequency contents of the cavitation signals excited at (a) 0.1 MPa, (b) 0.8 MPa and (c) 2 MPa PNP in Phantom 1, respectively, and (d-f) the corresponding spectrum of the cavitation signal. The PAM images were normalized by the maximum value in all PAM images.}
\label{fig:FrequencySelection}
\end{figure*}

To test whether AS-TAX can suppress artifacts while visualizing cavitation activity of different status, we used the harmonic, ultra-harmonic and broadband components (128 frequency bins) to reconstruct the image for Phantom 1, respectively (Fig.~\ref{fig:FrequencySelection}).
The comparison showed that the NNCC matrices calculated with the selected frequency components correctly reflected the spectral characteristic in the cavitation signals of different acoustic pressures (0.1 MPa, 0.8 MPa and 2 MPa) (Fig.~\ref{fig:FrequencySelection}(d)), akin to the AS reconstructed PAM images.
Consistent with previous observations, the TAX approach suppressed most of the X-shape artifacts in the NNCC matrix as compared with Fig.~\ref{fig:DAXTAXCoef}(d, f).
Therefore, the artifacts were well suppressed in the images reconstructed with a subset of the frequency components (Fig.~\ref{fig:FrequencySelection}(a--c)). 
Particularly, when excited at 0.1 MPa, the AS-TAX improved the ISNR by 12.1 dB compared to the AS method, which significantly improved the image quality for cavitation source localization when the bubble resonances are weak (Fig.~\ref{fig:FrequencySelection}(a)).

\subsection{Computational Cost}
\label{sec:ComputationalCost}
The computational complexity of the important operations of different methods is summarized in Table~\ref{table:ComputationalComplexity}, where $n_e$ is the number of transducer elements, $n_t$ is the number of time samples, $n_{\omega}$ is the number of selected frequency bins, $n_x$ and $n_z$ are the image size in the lateral and depth directions, respectively.
The DAX (TAX) operations in the time and frequency domain are comparable, whose complexities are $O(n_t n_z n_x)$ and $O(n_\omega n_z n_x)$, respectively.
Compared to the number of operations for image reconstruction with the AS and TEA methods, the complexity of the DAX (TAX) operation is negligible.
The most time-consuming operation in the AS method is the “IFFT” over the lateral
direction to obtain the map for each frequency, whose complexity is $O(n_\omega n_z n_x \log_2(n_x))$. 
Compared to the most time-consuming operations, the “Delay” and “Summation” ($O(n_e n_t n_z n_x)$) in the TEA method, this operation is two orders of magnitude faster with current settings.

\begin{table}
\centering
\caption{Computational complexity of the AS, TEA methods and DAX (TAX) operations.}
\label{table:ComputationalComplexity}
\setlength{\tabcolsep}{3pt}
\begin{tabular}{c|c|c}
\hline\hline
Method & Operation & {\makecell[c]{Computational complexity}} \\
\hline
\multirow{4}{*}{\makecell[c]{AS}} & {\makecell[c]{2D FFT}} & $O(n_x n_t \log_2(n_x n_t))$  \\
\multirow{4}{*}{} & Extrapolation
& $O(n_\omega n_z n_x)$ \\
\multirow{4}{*}{} & {\makecell[c]{IFFT}}
& $O(n_\omega n_z n_x \log_2(n_x))$  \\
\multirow{4}{*}{} & Summation over $\omega$
& $O(n_\omega n_z n_x)$ \\
\hline
\multirow{2}{*}{\makecell[c]{TEA}} & DAS
& $O(n_e n_t n_z n_x)$ \\
\multirow{2}{*}{} & Summation over $t$
& $O(n_e n_t n_z n_x)$ \\
\hline
\makecell[c]{DAX (TAX) \\ time domain} & Correlation
& \makecell[c]{$O(n_t n_z n_x)$}  \\
\hline
\makecell[c]{DAX (TAX) \\ frequency domain} & Correlation
& \makecell[c]{$O(n_\omega n_z n_x)$} \\
\hline\hline
\end{tabular}
\end{table}

The specific values for image reconstruction to compare the computational time required by the different methods for PAM were $n_e = 128$, $n_t = 2048$, $n_{\omega} = 128$ or $690$, $n_x = 256$, $n_z = 300$.
The average pre-run time for the delays in the TEA-based methods was 0.2 s. In contrast, the average pre-run time for establishing the spectral propagator for the AS-based methods with 128 and 690 frequency bins was 0.3 s and 2.4 s, respectively.
When using the same working bandwidth ($n_{\omega} = 690$) as the TEA-based methods, the run time of AS was 167 times faster than the TEA method (1.8 s vs 300.3 s) on the CPU.
The TEA-DAX added 26.5 s to the computational time required by the TEA (300.3 s) to reconstruct an image (Table~\ref{table:ComputationalTime}).
In contrast, the AS-DAX only added 2.5 s to the time required by the AS (1.8 s) to reconstruct an image.
As applying DAX and TAX to the AS method requires propagating the wave field for multiple times, the average run time was increased from 1.8 s (AS) to 4.3 s (AS-DAX) and then to 8.4 s (AS-TAX).
Nevertheless, the AS-TAX was still about 36 times faster than the TEA method.
When only using the harmonic, ultra-harmonic or broadband components ($n_{\omega} = 128$) for image reconstruction, the AS-TAX was about 214 times faster than the TEA method.

On the GPU, the added computational time of TEA-DAX compared to that of TEA was negligible (16.9 ms vs 17.2 ms) (Table~\ref{table:ComputationalTime}).
For using 690 frequency bins, the average run time of the AS, AS-DAX and AS-TAX methods were 3.6 ms, 6.7 ms and 9.8 ms, respectively (Table~\ref{table:ComputationalTime}).
This shows that the added computational time for applying DAX and TAX to the AS method was associated with the time to form an image with AS on the GPU, which is consistent with the observations on the CPU.
Nevertheless, the AS-TAX was still about 1.7 times faster than the TEA method on the GPU. 
For using 128 frequency bins, the average run time of the AS, AS-DAX and AS-TAX methods were 0.6 ms, 1.2 ms and 1.8 ms, respectively (Table~\ref{table:ComputationalTime}) and the AS-TAX was about 9.4 times faster than the TEA method on the GPU.

\begin{table}
\centering
\caption{Computational time (mean $\pm$ standard deviation) of the AS, AS-DAX, AS-TAX, TEA and TEA-DAX methods to reconstruct an image in the phantom experiments.}
\label{table:ComputationalTime} 
\setlength{\tabcolsep}{3pt}
\begin{tabular}{c|c|c|c|c}
\hline\hline
Method & 
{\makecell[c]{$n_{\omega}$ \\ or $n_t$}} & {\makecell[c]{Pre-run time \\ on CPU (s)}} & {\makecell[c]{Run time \\ on CPU (s)}} & {\makecell[c]{Run time \\ on GPU (ms)}} \\
\hline
\multirow{2}{*}{AS} & 690 & 2.4 $\pm$ 0.1 & 1.8 $\pm$ 0.1 & 3.6 $\pm$ 0.2 \\
\multirow{2}{*}{} & 128 & 0.3 $\pm$ 0.1 & 0.4 $\pm$ 0.01 & 0.6 $\pm$ 0.06 \\
\hline
\multirow{2}{*}{AS-DAX} & 690 & 2.4 $\pm$ 0.1 & 4.3 $\pm$ 0.1 & 6.7 $\pm$ 0.3 \\
\multirow{2}{*}{} & 128 & 0.3 $\pm$ 0.1 & 0.6 $\pm$ 0.04 & 1.2 $\pm$ 0.1 \\
\hline
\multirow{2}{*}{AS-TAX} & 690 & 2.4 $\pm$ 0.1 & 8.4 $\pm$ 0.1 & 9.8 $\pm$ 0.4 \\
\multirow{2}{*}{} & 128 & 0.3 $\pm$ 0.1 & 1.4 $\pm$ 0.1 & 1.8 $\pm$ 0.1 \\
\hline
TEA & 2048 & 0.2 $\pm$ 0.02 & 300.3 $\pm$ 1.1 & 16.9 $\pm$ 0.6 \\
\hline
TEA-DAX & 2048 & 0.2 $\pm$ 0.02 & 326.8 $\pm$ 4.7 & 17.2 $\pm$ 0.8 \\
\hline\hline
\end{tabular}
\end{table}

In terms of memory consumption, the TEA-based and AS-based methods required about 38 MB memory and 75 MB (404 MB) to storage the pre-calculated delays and spectral propagators for 128 (690) frequency bins, respectively. 
On the GPU, the memory required by the TEA-based methods was 600 MB, while AS, AS-DAX and AS-TAX required 75 MB (404 MB), 150 MB (809 MB) and 225 MB (1213 MB) for 128 (690) frequency bins, respectively.


\section{Discussion}
\label{sec:Discussion}
The capability of monitoring cavitation activity with high-resolution is important for expediting the clinical adoption of cavitation-based therapies~\cite{Schoen2022,lee2022}.
Toward the goal of accurately localizing stable and inertial cavitation in the space and time, we developed the AS-TAX method, an extension of the DAX approach to the frequency domain.
We showed that 1) cross-correlating two time-varying fields in the time and frequency domain are equivalent; 2) taking a portion of the frequency components of the wave fields for cross-correlation can effectively suppress artifacts in the images associated with the same frequency components; 3) the pattern of the NNCC matrix can be varied spatially by manipulating the apodization functions, to avoid the X-shape artifacts induced by the diffraction pattern of diagnostic imaging array~\cite{Lu2019,haworth2012passive}.
With these findings, we presented the AS-TAX method as a fast image reconstruction method with high image SNR and with the capability to localize stable and inertial cavitation simultaneously, which is important for reducing risks of damaging normal tissues and improving therapeutic outcomes in cavitation-based therapies~\cite{JI2021458,HU2023106346}.

The artifacts suppression capability of DAX and TAX depends on the selection of the apodization functions.
In previous works for PAM~\cite{Lu2019,Lu2020}, the pattern of the apodization functions was selected based on evaluating the improved image SNR.
In this work, we further included the centroid in the NNCC matrix indicating source localization accuracy to determine the pattern, based on the rationale that the NNCC value should have the maximum in the main lobe as well.
Using the phantom data, we arrived at the optimal pattern of the apodization functions (Pattern-16) for an array of 128 elements for our experiments.
This pattern also works well in the mouse tumor data for artifacts suppression (Fig.~\ref{fig:ExperimentalResultsDifferentMethod}(c)).
Whether the optimal pattern depends on the parameters of the ultrasound arrays (elements number, pitch, working bandwidth, etc.) remains further study.
It is noted that the active and inactive elements in the complementary apodization functions can be randomly chosen (random pattern)~\cite{seo2008sidelobe}.
However, consistent with previous findings~\cite{Lu2019,Lu2020}, we also observed that this pattern does not remarkably improve the image quality in PAM (data not shown).
While the aperture can also be subdivided using quatra or octa apodization, the increased number of sub-apertures will introduce additional computational cost to compute the NCC matrix and we empirically observed that the weaker source is often missed when monitoring multiple acoustic sources with too many sub-apertures.

The results consistently showed that the AS-TAX method (as compared to TEA) reduced the ESA (by 34.8 -- 66.7\%) and improved ISNR (by 10.7 -- 14.5 dB) across various imaging depth for various numbers of cavitation sources, in the phantom and animal experiments (Table~\ref{table:SimulatedAndExperimentalPerformance}).
The ESA was smaller in the mouse images (source depth at 27 mm) than the ones in Phantom 1 (source depth at 40 mm) which can be attributed the fact that the size of the point-spread-function in PAM is larger as the imaging depth goes deeper~\cite{haworth2016quantitative,gray2020}.
Meanwhile, both the images and the metrics in all the experiments showed that AS-DAX and TEA-DAX are equivalent in terms of artifacts suppression, evidencing the fact that the NCCs can be calculated in the time or frequency domain (Appendix\ref{sec:CoefficientDerivation}).
Thus, DAX and TAX can actually be combined with the TEA method both in the time domain and frequency domain (Appendix\ref{sec:FDTEA}), as well as with the helical wave spectrum method for PAM with convex arrays~\cite{Zhu2024PAM}.

The recently proposed data-adaptive spatial filtering approach for PAM~\cite{haworth2023passive} suppresses artifacts significantly with low computational cost.
However, accurate estimation of the required threshold for the data-adaptive spatial filter in experiments is difficult.
In contrast, the AS-TAX is relatively easy for obtaining the optimal patterns for effective artifacts removal.
Compared to other data-adaptive beamformers (Appendix\ref{sec:ComparisionAdaptive}), the AS-TAX achieved comparable artifacts suppression results with a much lower computational cost.
In the meantime, the AS-TAX maintains well the relative cavitation strength ratio
between multiple cavitation sources under the influence of noise (Appendix\ref{sec:SimulatedResults}), which may be important for detecting off-target cavitation events~\cite{song2011investigation}.
Nevertheless, one must notice that the DAX and TAX schemes do not improve the spatial resolution in PAM (Appendix\ref{sec:SpatialResolution}).
This is because the wave fields reconstructed by different sub-apertures are highly correlated in the main-lobe, thus applying the NCC matrix to the images does not alter the shape of the main-lobe~\cite{haworth2016quantitative}. 

The NNCC matrix calculated with the cavitation status informed frequency components (i.e., harmonic, ultra-harmonic and broadband components) in AS-TAX (Fig.~\ref{fig:FrequencySelection}), evidenced that the wave fields propagated by AS for different apodized apertures were correlated at these different frequencies.
After all, when the signal energy of these frequency components are contributed by the MB cavitation activity, they are from the same source location and the back-propagated wave fields should be correlated at these frequencies. 
Another important evidence is that,  when the MBs cavitated at low pressure (0.1 MPa) without emitting ultra-harmonics and broadband cavitation signals (Fig.~\ref{fig:FrequencySelection}(a, d)), high correlation values were not presented at the cavitation source location in the NNCC matrix formed using these frequencies.
The remarkably improved ISNR by AS-TAX (12.1 dB compared to AS) for very small amplitude of cavitation (Fig.~\ref{fig:FrequencySelection}(a)) also demonstrated its potential in the applications where relatively weak cavitation strength must be maintained, e.g. for gene delivery to stem cells, as cell viability is a key factor for successful treatments~\cite{bez2017}.

Parallel implementation of TEA on GPUs has been employed in many of the previous works~\cite{collin2013real,o2014three,jones2015experimental,jones2018three} to realize real-time image reconstruction speed, because the DAS and summation operations in TEA are independent between pixels, which are particularly suitable for parallelization on GPUs.
In addition, the DAX and TAX operations are independent between pixels, as well.
Thus, the advantage of the AS-based methods over the TEA-based methods in terms of computational time was reduced when deployed on the GPU compared to on the CPU, although the computational complexity of the former was lower (Table~\ref{table:ComputationalTime}).
Yet, the AS-based methods were still faster than their TEA-based and FD-TEA based counterparts either on CPU or on GPU (Table~\ref{table:ComputationalTime}, Appendix\ref{sec:FDTEA}).
While the run time of TEA can be improved significantly by sparse matrix multiplication~\cite{kamimura2020real}, it requires several hours to build the sparse matrix (pre-run time) which is scaled by the image size and the number of time samples in the RF signals, and has to be carried out offline.
When the settings need to be updated frequently, such as when dealing with heterogeneous medium, the practical value of the sparse matrix multiplication approach is limited. 
In contrast, AS-TAX only requires a few seconds (pre-run time) to set up the spectral propagator (Table~\ref{table:ComputationalTime}).
Thus, it is more flexible than the TEA with sparse matrix approach as the necessary operators can be calculated on-the-fly.

The added time when AS is combined with the DAX and TAX is related to how many times the wave field needs to be back propagated, i.e., how many sub-apertures are used to group the cavitation signals.
Therefore, the differences between AS and AS-DAX, and between AS-DAX and AS-TAX were basically the time used by AS for wave field extrapolation (Eq.~\eqref{PropagateAS}), which was about 3 ms on the GPU with the current settings.
In terms of memory consumption, AS-DAX and AS-TAX require twice and three times the memory compared to AS, respectively, which is also linked to how many times the wave field needs to be back propagated.
Note also that the computational time in the AS-based methods is closely related to the number of selected frequency bins, because the image is superimposed by maps created for different frequencies.
When imaging either stable or inertial cavitation (using only 128 bins), the acceleration (9.4 times) made by AS-TAX and its lower memory consumption compared to TEA on the GPU is particularly appealing.

These results showed that the AS-TAX method represents an efficient and precise way to visualize the spatial distribution of different types of cavitation activities with high ISNR and can be used to form a closed-loop controller for maintaining target levels of stable and inertial cavitation in the space and time, such as the one demonstrated in~\cite{patel2018closed}.

 
\section{Conclusion}
In this work, we extended the DAX approach for PAM to the frequency domain to be combined with the AS method for artifacts suppression with improved image reconstruction speed. We further developed the TAX approach to suppress the X-shape artifacts existed in the DAX-based methods. The results demonstrated that AS-TAX has comparable image reconstruction quality as the data-adaptive beamformers, which remarkably reduced the ESA and improved the ISNR compared to TEA while being faster than TEA both on CPU and GPU. Furthermore, AS-TAX can reconstruct high-quality PAM images using selected frequency components for visualizing cavitation activities of different status. Therefore, the proposed method is promising in real-time monitoring and accurate localization of MB cavitation activity in the applications of FUS therapy.
\appendices

\section*{Appendix}
\subsection{Derivation of frequency domain calculation of NCCs}
\label{sec:CoefficientDerivation}

Here, we show that the NCCs calculated in the time domain by \eqref{CorrelationCoeff} and the frequency domain by \eqref{CorrelationCoeffAS} are equivalent. Considering two beamformed signals $\boldsymbol{x}$ and $\boldsymbol{y}$ having 0 offsets in the time domain, the NCC is given by:
\begin{equation}
\begin{aligned}
\rho_{TD} = \frac{\boldsymbol{x}^T\boldsymbol{y}}{\sqrt{\boldsymbol{x}^T\boldsymbol{x}}\sqrt{\boldsymbol{y}^T\boldsymbol{y}}}
\label{CoefficientTD2}
\end{aligned}
\end{equation}
In the frequency domain, the NCC is given by:
\begin{equation}
\begin{aligned}
\rho_{FD} = \frac{\boldsymbol{X}^H\boldsymbol{Y}}{\sqrt{\boldsymbol{X}^H\boldsymbol{X}}\sqrt{\boldsymbol{Y}^H\boldsymbol{Y}}}
\label{CoefficientFD1}
\end{aligned}
\end{equation}
with $\boldsymbol{X}$ and $\boldsymbol{Y}$ the Fourier transform of $\boldsymbol{x}$ and $\boldsymbol{y}$, respectively.

The Wiener-Khinchin theorem\cite{chatfield2013analysis} states that the power spectral density $|X(\omega)|^2=X(\omega)^*X(\omega)$ of a signal $x(t)$ corresponds to the Fourier transform of its auto-correlation function $R_{xx}(\tau)$, thus:
\begin{equation}
\begin{aligned}
R_{xx}(\tau) &= \int_{-\infty}^{\infty} x^*(t)x(t-\tau)dt \\
&= \frac{1}{2\pi}\int_{-\infty}^{\infty} X^*(\omega)X(\omega)e^{i\omega\tau}d\omega
\label{Wiener-Khinchin1}
\end{aligned}
\end{equation}
Where $(.)^*$ is the conjugate operation.
This conclusion can be generalized to cross-correlation \cite{smith2008mathematics}:
\begin{equation}
\begin{aligned}
R_{xy}(\tau) &= \int_{-\infty}^{\infty} x^*(t)y(t-\tau)dt \\
&= \frac{1}{2\pi}\int_{-\infty}^{\infty} X^*(\omega)Y(\omega)e^{i\omega\tau}d\omega
\label{Wiener-Khinchin2}
\end{aligned}
\end{equation}
With Eqs. \eqref{Wiener-Khinchin1}-\eqref{Wiener-Khinchin2}, Eqs. \eqref{CoefficientTD2}-\eqref{CoefficientFD1} can be reformulated as:

\begin{equation}
\begin{aligned}
\rho_{TD} &= \frac{\int_{-\infty}^{\infty} x^*(t)y(t)dt}{\sqrt{\int_{-\infty}^{\infty} x^*(t)x(t)dt}\sqrt{\int_{-\infty}^{\infty} y^*(t)y(t)dt}} \\
&= \frac{R_{xy}(0)}{\sqrt{R_{xx}(0)}\sqrt{R_{yy}(0)}}
\label{CoefficientTD3}
\end{aligned}
\end{equation}

\begin{equation}
\begin{aligned}
\rho_{FD} &= \frac{\int_{-\infty}^{\infty} X^*(\omega)Y(\omega)d\omega}{\sqrt{\int_{-\infty}^{\infty} X(\omega)X^*(\omega)d\omega}\sqrt{\int_{-\infty}^{\infty} Y^*(\omega)Y(\omega)d\omega}} \\
&= \frac{R_{xy}(0)}{\sqrt{R_{xx}(0)}\sqrt{R_{yy}(0)}}
\label{CoefficientFD3}
\end{aligned}
\end{equation}
Therefore, $\rho_{TD}$ and $\rho_{FD}$ are equivalent. In frequency domain, the DAX methods calculate the NCC using only positive frequency components, resulting in complex numbers in the NCCs. To avoid this, we generate the negative frequency components as the conjugate of the positive frequency components. Subsequently, the positive and negative frequency components were combined to calculate the NCC.

\subsection{Comparison with other PAM beamformers}
\label{sec:Comparision}
\subsubsection{PAM with DAX and TAX}
\label{sec:FDTEA}
\begin{figure}
\centerline{\includegraphics[width=0.8\columnwidth]{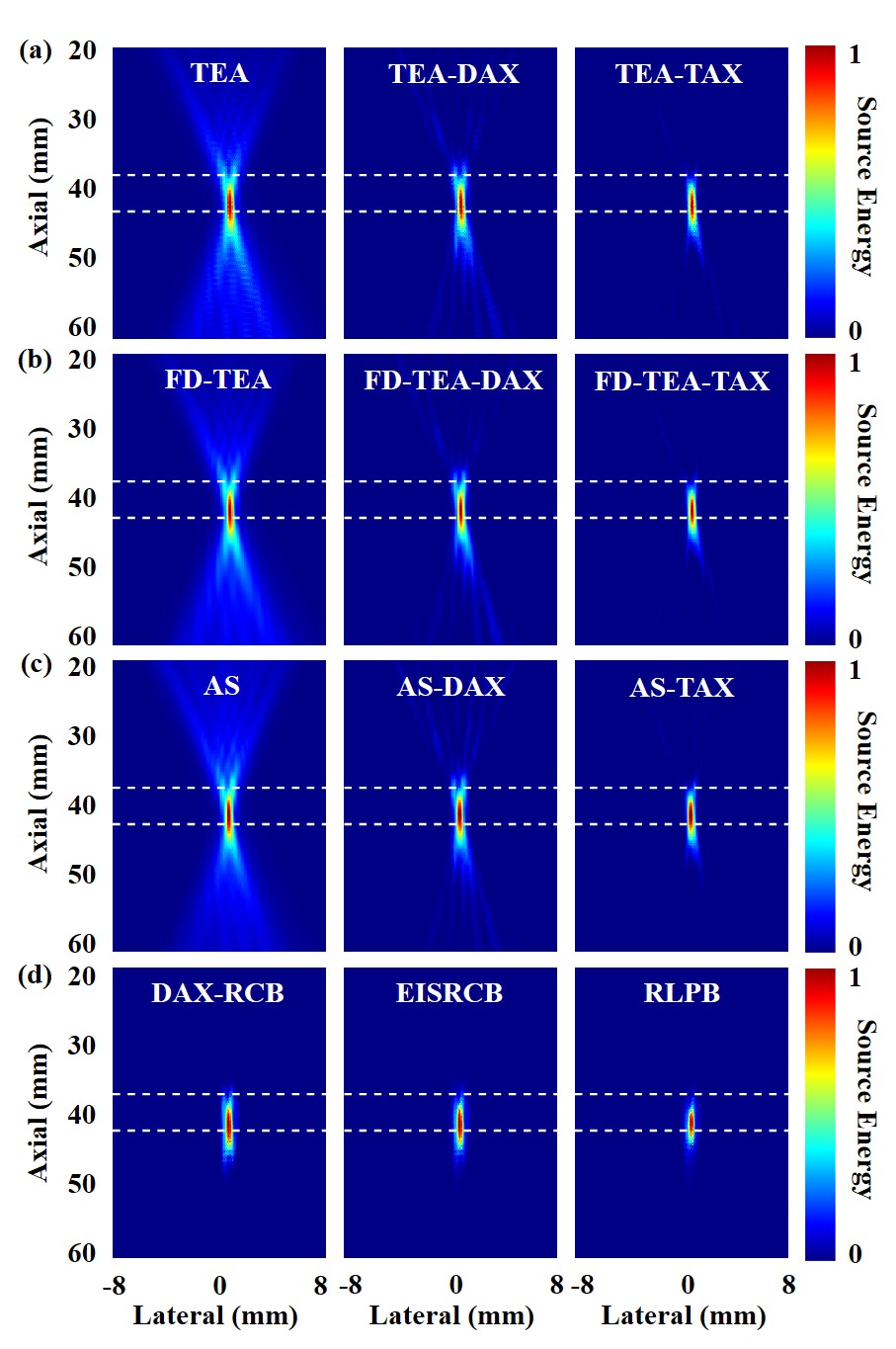}}
\caption{PAM images reconstructed by (a) TEA, TEA-DAX, TEA-TAX, (b) FD-TEA, FD-TEA-DAX, FD-TEA-TAX, (c) AS, AS-DAX, AS-TAX, (d) DAX-RCB, EISRCB and RLPB for Phantom 1. The white dash lines outline the tunnel's boundary.}
\label{fig:CompareResults}
\end{figure}

Like TEA-DAX, TAX can be combined with TEA which becomes the TEA-TAX method.
In addition, TEA can be performed in frequency domain (FD-TEA) with the harmonic wave field at $\boldsymbol{r}$ obtained following~\cite{haworth2016quantitative}:
\begin{equation}
\tilde{p}(\boldsymbol{r},\omega)=\sum_{l=1}^{L} \mathcal{F}_t\{s_l(t)\}e^{i\omega\tau_l(\boldsymbol{r})}
\label{DelayedSignalFD}
\end{equation}
Thus, FD-TEA can be also combined with DAX and TAX for artifacts suppression.
Here, we extend previous comparison in Fig.~\ref{fig:ExperimentalResultsDifferentMethod}(a), to demonstrate the equivalence of applying the DAX and TAX schemes to the wave field obtained with TEA, FD-TEA and AS in terms of image quality (Fig.~\ref{fig:CompareResults}, Table.~\ref{table:DiffenrenMethodPerformance}).
For the comparison, the wideband frequency components were used for image reconstruction.

The differences in the averaged ESA and ISNR between TEA-DAX, FD-TEA-DAX and AS-DAX were less than 0.29 $mm^2$ and 0.46 dB, respectively.
For the TAX-based methods, the differences were less than 0.23 $mm^2$ and 0.82 dB, respectively.
The improvements in the image quality with the DAX and TAX schemes were more prominent in ISNR compared to their counterparts, which was between 6.16 -- 11.24 dB.
In terms of the computational cost, the AS-based methods were faster than their counterparts with both TEA and FD-TEA on CPU and GPU (see also the comparison in Table.~\ref{table:ComputationalTime}).
The averaged run time of TEA-TAX, FD-TEA, FD-TEA-DAX and FD-TEA-TAX on the CPU (GPU) were 435.8 s (19.2 ms), 85.3 s (14.3 ms), 105.6 s (15.3 ms) and 107.1 s (17.3 ms), respectively.

\begin{table}
\centering
\caption{ESA and ISNR (mean $\pm$ standard deviation) of the TEA, TEA-DAX, TEA-TAX, FD-TEA, FD-DAX, FD-TAX, AS, AS-DAX and AS-TAX methods for the Phantom 1.}
\label{table:DiffenrenMethodPerformance}
\setlength{\tabcolsep}{5pt}
\begin{tabular}{c|c|c}
\hline\hline
Algorithm & {\makecell[c]{ESA ($mm^2$)}} & {\makecell[c]{ISNR (dB)}} \\
\hline
TEA & 5.94 $\pm$ 2.56 & 12.23 $\pm$ 0.79 \\
FD-TEA & 6.16 $\pm$ 2.76 & 12.34 $\pm$ 0.79 \\
AS & 4.81 $\pm$ 2.22 & 11.65 $\pm$ 0.84 \\
\hline
TEA-DAX & 3.77 $\pm$ 1.34 & 18.50 $\pm$ 0.75 \\
FD-TEA-DAX & 3.63 $\pm$ 1.38 & 18.80 $\pm$ 0.85 \\
AS-DAX & 3.48 $\pm$ 1.35 & 18.96 $\pm$ 0.75 \\
\hline
TEA-TAX & 2.30 $\pm$ 0.91 & 22.07 $\pm$ 1.16 \\
FD-TEA-TAX & 2.37 $\pm$ 0.94 & 22.42 $\pm$ 1.12 \\
AS-TAX & 2.14 $\pm$ 0.76 & 22.89 $\pm$ 1.06 \\
\hline
DAX-RCB & 2.44 $\pm$ 0.94 & 23.95 $\pm$ 1.05 \\
EISRCB & 1.90 $\pm$ 0.99 & 26.09 $\pm$ 1.41 \\
RLPB & 1.50 $\pm$ 0.79 & 23.13 $\pm$ 1.18 \\
\hline\hline
\end{tabular}
\end{table}

\subsubsection{AS-TAX versus data-adaptive beamformers}
\label{sec:ComparisionAdaptive}
We also implemented the data-adaptive beamformers for PAM including DAX-RCB~\cite{Lu2020}, EISRCB~\cite{Lu2018} and RLPB~\cite{Lyka2018} and evaluated the performance of AS-TAX and the data-adaptive beamformers with the data reported in Fig.~\ref{fig:ExperimentalResultsDifferentMethod}(a) for comparison. 
The cavitation source was localized (average value) at (-0.59 mm, 40.61 mm), (-0.68 mm, 40.92 mm), (-0.70 mm, 40.79 mm) and (-0.69 mm, 40.73 mm) in the images reconstructed by AS-TAX, DAX-RCB, EISRCB and RLPB, respectively.
The differences in the average ESA and ISNR between the AS-TAX and the adaptive beamformers were less than 0.64 $mm^2$ and 3.74 dB (Table.~\ref{table:DiffenrenMethodPerformance}).
However, AS-TAX is remarkably computationally efficient than the data-adaptive beamformers.
The average computational time required by DAX-RCB, EISRCB and RLPB to reconstruct one image of 300 $\times$ 256 pixels was 675 s, 697 s and 13824 s using a single CPU thread, while it was 1.4 s with AS-TAX (about 80--1646 times faster).
These results demonstrated that the AS-TAX has comparable image reconstruction quality as the data-adaptation beamformers, while its computational cost is obviously much lower.

\subsection{Further evaluation of the AS-based methods}
\label{sec:FurtherEvaluation}
\subsubsection{Spatial resolution} 
\label{sec:SpatialResolution}
\begin{figure}
\centerline{\includegraphics[width=1.0\columnwidth]{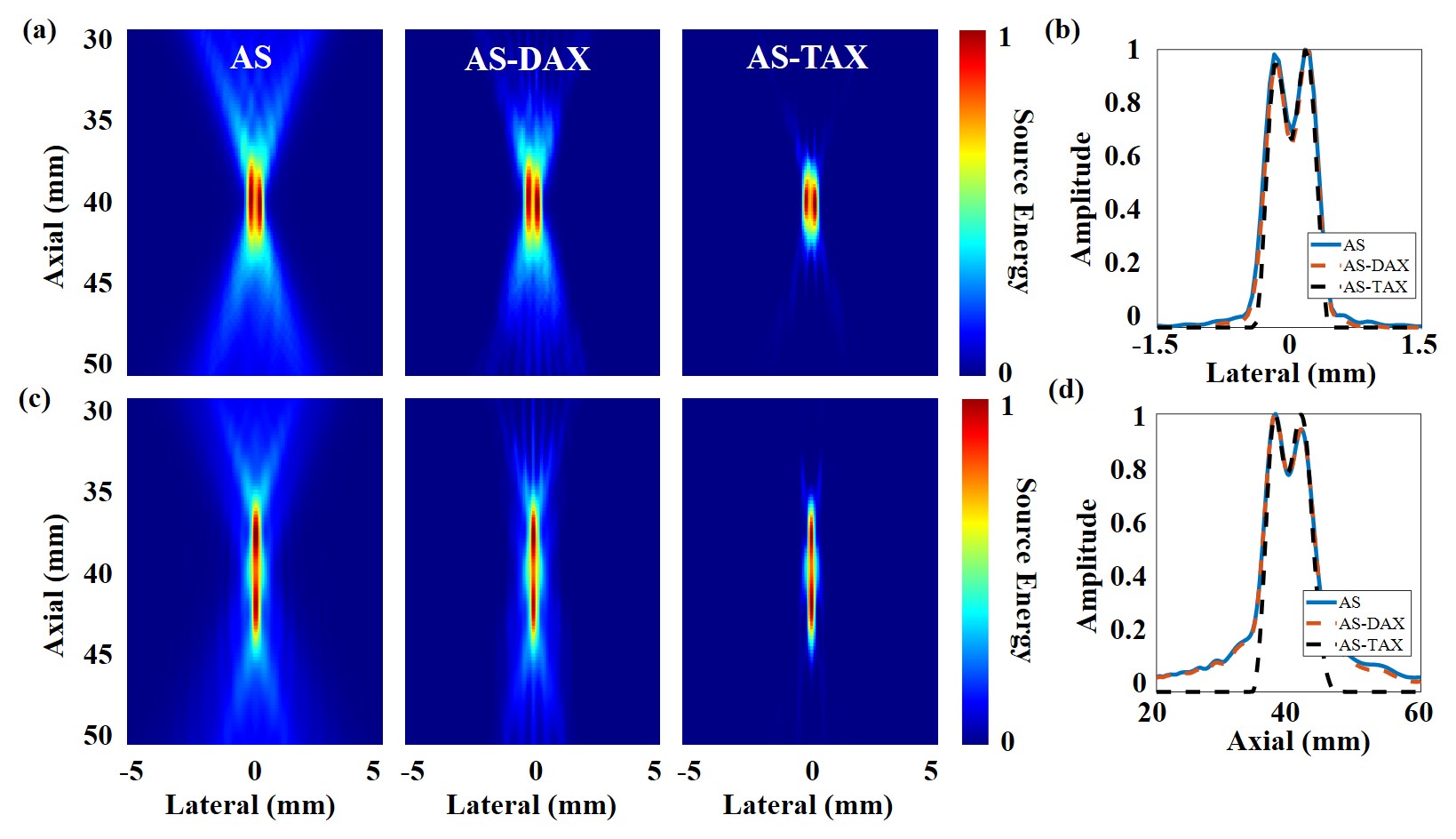}}
\caption{Simulated PAM images reconstructed by AS, AS-DAX and AS-TAX for two closely distributed bubbles (a) at the same depth and (b) the corresponding lateral profile crossing the two bubbles, (c) in the center of the array laterally and (d) the corresponding axial profile crossing the two bubbles.}
\label{fig:SpatialResolution}
\end{figure}

According to Rayleigh’s criterion\cite{michalet2006,Lyka2018}, the height of the trough between two point-targets in the image should be 26.3\% less than that of the peak to resolve them as two distinct objects.
We, therefore, simulated the cavitation signals of two closely distributed bubbles using the Marmonttant model\cite{Marmottant2005} to evaluate the spatial resolution of AS, AS-DAX and AS-TAX.
The cavitation signals from two bubbles with diameter of 1 $\mu$m and 1.1 $\mu$m, respectively, were excited by 1 MHz excitation signals (100 cycles) at 250 kPa.
For evaluating the lateral/axial resolution, the bubbles were positioned at (-0.175 mm, 40 mm)/(0.175 mm, 40 mm) and (0 mm, 38 mm)/(0 mm, 42 mm), respectively (Fig.~\ref{fig:SpatialResolution}(a) -- (c)).
To receive the bubbles' emissions in a homogeneous medium ($c=1500$ m/s), the CL15-7 array (8.9 MHz, 67\% bandwidth) with 25.6 mm aperture size was simulated using k-Wave\cite{Bradley2010}. The received signals (2048 time samples) were sampled at 35.6 MHz.

For the two bubbles located at the same depth ($z=40$ mm), the trough of the lateral profile in the AS, AS-DAX and AS-TAX images were 30\%, 35\% and 33\%, respectively (Fig.~\ref{fig:SpatialResolution}(b)). 
For the two bubbles distributed in the center of the array ($x=0$ mm), the trough of the axial profile in AS, AS-DAX and AS-TAX was 27\%, 28\% and 28\%, respectively (Fig.~\ref{fig:SpatialResolution}(d)). Thus, the three methods have similar spatial resolution as evaluated by the Rayleigh’s criterion.
 
\subsubsection{Noise and uneven cavitation strength} 
\label{sec:SimulatedResults}
One may wonder whether a spatial mask defined by the AS image itself can improve the image quality if applied directly just like the NCC matrix used in the AS-TAX method.
To test this hypothesis, we define the AS$^3$ method with the following equation:
\begin{equation}
\begin{aligned}
I(\boldsymbol{r}) = (I_{AS}(\boldsymbol{r})/Max(I_{AS}(\boldsymbol{r}))^2 I_{AS}(\boldsymbol{r})
\label{AS3}
\end{aligned}
\end{equation}
Where $I_{AS}$ is the AS image calculated using Eq.~(\ref{PAMAS}).
The AS$^3$ method can be regarded as using the squared AS image itself instead of using the $\rho_n$ in Eq.~(\ref{EnergyASTAX}) to suppress artifacts.
We then compared the AS-TAX and AS$^3$ method using the simulated cavitation signals received by the CL15-7 array for imaging the two bubbles (diameter: 1.1$\mu$m) using the Marmonttant model and the settings as described in Appendix\ref{sec:SpatialResolution}. The two bubbles were positioned at (-5 mm, 40 mm) and (5 mm, 40 mm), respectively.
Cavitation signals emitted from different bubbles were set to having the same phase but different amplitudes (1:2), resulting in uneven cavitation strength (1:4). 
White noise was added to the received signals using the Matlab function \texttt{awgn}, to adjust the SNR to 15 dB, 10 dB and -5 dB, respectively.

\begin{figure}
\centerline{\includegraphics[width=0.8\columnwidth]{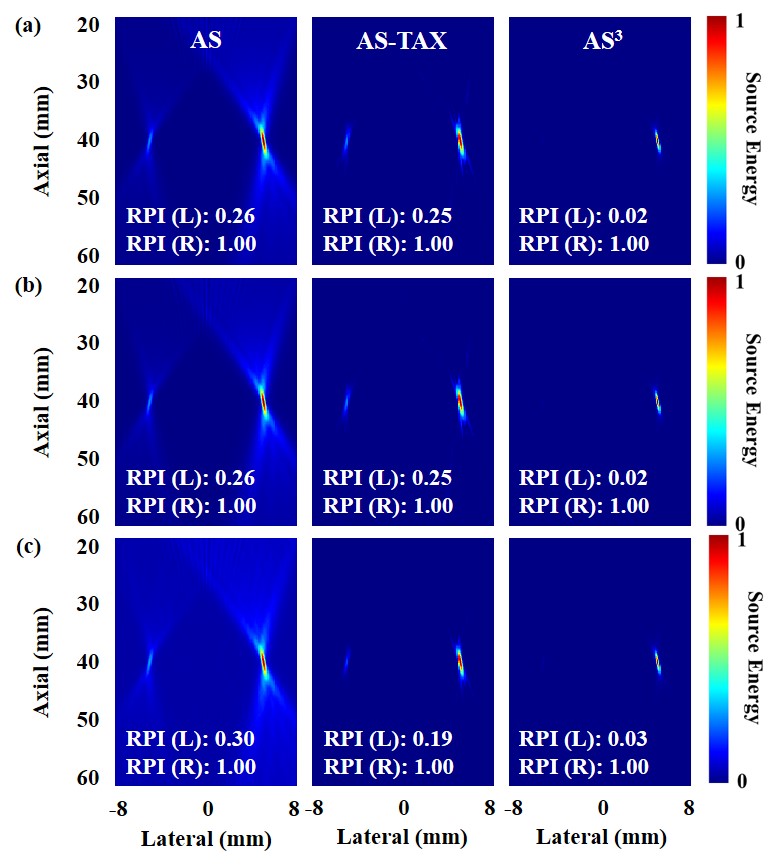}}
\caption{PAM images reconstructed by AS, AS-TAX and AS$^3$ for the simulated multiple cavitation sources with uneven strength of (a) 15 dB, (b) 5 dB and (c) -5 dB SNR. RPI (L): RPI of the left source; RPI (R): RPI of the right source.}
\label{fig:SimulatedResults}
\end{figure}

To evaluate the alterations on the imaged source-strength by the different methods, the PAM images reconstructed by the AS-TAX and AS$^3$ method were normalized to the maximum intensity in the AS-beamformed image for each noise level (Fig.~\ref{fig:SimulatedResults}).
The regional peak intensity (RPI), i.e, the local maximal intensity value, of each source is then defined for the comparison.
With AS$^3$, the weaker cavitation source (left bubble) was hardly visible as the values in the spatial mask defined by the AS image for the left bubble were less than 0.07 which scales down the estimated source strength.
In fact, when imaging multiple cavitation sources with uneven strength, the AS$^3$ may not be a good choice for artifacts suppression, because the non-linear scaling applied by the spatial mask always tends to weaken the cavitation strength of the weaker source, from 0.26-0.30 to 0.02-0.03 in RPI.
With AS-TAX, this issue can be avoided (Fig.~\ref{fig:SimulatedResults}).
In our data, when the SNR is greater than 5 dB, the RPI of the weaker cavitation source with AS-TAX was 0.25 versus 0.26 with AS.
At -5 dB SNR, the RPI of the weaker cavitation with AS-TAX was altered a bit, 0.19 versus 0.30 with AS, but was still better than the value with AS$^3$ (0.03).
Thus, the AS-TAX well preserved the relative cavitation strength ratio between multiple cavitation sources which may be important for detecting off-target cavitation events.


\bibliographystyle{IEEEtran}
\bibliography{IEEEabrv,bibfiles}

\end{document}